\documentclass[10pt,preprint]{aastex}

\usepackage{color}
\usepackage{natbib}

\begin{document}

\title{Improving interferometric null depth measurements using statistical distributions: theory and first results with the Palomar Fiber Nuller}


\author{C. Hanot\altaffilmark{1}, B. Mennesson\altaffilmark{2}, S. Martin\altaffilmark{2}, K. Liewer\altaffilmark{2}, F. Loya\altaffilmark{2}, D. Mawet\altaffilmark{2}, P.~Riaud\altaffilmark{1}, O. Absil\altaffilmark{1} and E.~Serabyn\altaffilmark{2}  }
\affil{$^1$Institut d'Astrophysique et de G\'eophysique, University of Li\`ege, \\ All\'ee du 6 Ao\^ut, 17 B\^at B5c, 4000 Li\`ege, Belgium}
\affil{$^2$Jet Propulsion Laboratory, California Institute of Technology, \\ 4800 Oak Grove Drive, Pasadena,
California 91109, USA}
\email{hanot@astro.ulg.ac.be}

\begin{abstract} A new "self-calibrated" statistical analysis method  has been developed for the reduction of nulling interferometry data. The idea is to use the statistical distributions of the fluctuating null depth and beam intensities to retrieve the astrophysical null depth (or equivalently the object's visibility) in the presence of fast atmospheric fluctuations. The approach yields an accuracy much better (about an order of magnitude) than is presently possible with standard data reduction methods, because the astrophysical null depth accuracy is no longer limited by the magnitude of the instrumental phase and intensity errors but by uncertainties on their probability distributions. This approach was tested on the sky with the two-aperture fiber nulling instrument mounted on the Palomar Hale telescope. Using our new data analysis approach alone - and no observations of calibrators - , we find that error bars on the astrophysical null depth as low as a few $10^{-4}$ can be obtained in the near infrared, which means that null depths lower than $10^{-3}$ can be reliably measured.  This statistical analysis is not specific to our instrument and may be applicable to other interferometers.
\end{abstract}


\maketitle 
\section{Introduction}
Since the first discovery of an exoplanet around a solar-type star \citep{Mayor95}, the quest to find earth-like exoplanets and, even more importantly detect the presence of life on them became a major topic in astrophysics. However, the direct imaging of such systems is very challenging because of the high spatial resolution and dynamic range required. One possible way to overcome these difficulties is to use nulling interferometry \citep{Bracewell78}. In this approach, one destructively combines the light coming from two or more apertures in order to dim the bright on-axis starlight  and reveal faint objects or structures in the immediate vicinity.

The analysis of interferometric data in general, and nulling interferometric data in particular \citep{Colavita09}, is a complex task because accurate calibration of the instrument is needed to extract the scientific information. In the case of interferometric nulling, the quantity of interest is the astrophysical null depth ($N_a$), which is the inverse of the rejection ratio, and directly relates to the target's spatial brightness distribution. In practice however, the measured interferometric null depth is not strictly equal to the astrophysical null depth, because of the effects of instrumental noise and error sources such as phase differences, intensity mismatch and global intensity fluctuations.  It had been thought that a proper determination of the astrophysical null depth requires the mean values of these instrumental error sources to be accurately known, e.g. \citep{Serabyn00,Lay04}.  The classical method used for deriving astrophysical null depths - and visibilities - has therefore been to average different sequences recorded on the science star and estimate the instrumental bias by observing a calibrator star \citep{Colavita09}. This technique has been extensively used for years for both classical and nulling interferometry, but suffers from well known limitations: (i), the final accuracy depends on the scientific knowledge of the calibrator star, (ii), the accuracy is limited by the stability of the observing conditions   and (iii), calibrator observations are time consuming. 

To circumvent these limitations, we describe here a new method of calibrating astrophysical null depths, based on measuring the properties of the observed null depth {\it {distribution}}.  The basic idea is to record a time sequence of the rapid null depth fluctuations, and then retrieve the underlying astrophysical information by modeling the observed statistics of the null depth distribution. Using such a statistical analysis, we show in the following that it is possible to retrieve astrophysical null depths with  much better accuracy than classical approaches allow.  Moreover, our initial stellar observations indicate that this statistical approach does not require any observation of calibrator stars, at least down to null depth measurement accuracies as low as  a few $10^{-4}$ (the exact number  depends on the instrument set-up being used).  In this paper, we first explain the principle and theory of this new statistical data analysis strategy, and then apply it to initial astronomical null data obtained with the Palomar Fiber Nuller (PFN) \citep{Serabyn06a, Serabyn06b, Mennesson06,Martin08}, a nulling-based interferometric ``coronagraph'' developed at the Jet Propulsion Laboratory.  However, we emphasize that the new reduction method can potentially be applied to any null and/or visibility measurements in general.
\\
\section{The statistics of the null depth}

\subsection{The expression for the null}

We begin from the expression for the observed null depth of a two beam interferometer for a point source in the presence of error sources, as given by  \citet{Serabyn00}. In the case of two planar monochromatic wavefronts, the combined stellar intensity measured at constructive interference (+) and destructive interference (-) at time $t$ is given by :
\begin{eqnarray}
I_\pm^{*}(t) &=& \frac{1}{2}\left[ I_1^{*}(t)+I_2^{*}(t)\pm 2 \cos\left(\Delta\phi(t)\right)\cos\left(\alpha_{rot}\right)\sqrt{I_1^{*}(t)I_2^{*}(t)}\right] \\
&=&  \langle I^{*}(t) \rangle \left[ 1\pm\cos\left(\Delta\phi(t)\right)\cos\left(\alpha_{rot}\right)\sqrt{1-\left(\delta I(t)\right)^2}\right]
\label{intpm}
\end{eqnarray}
where $I_1^{*}(t)$ and $I_2^{*}(t)$ are the individual stellar intensities of beams 1 and 2 at the beam combiner, respectively, $ \langle I^{*}(t) \rangle =(I_1^{*}(t)+I_2^{*}(t))/2$ is the average input beam intensity, $\delta I(t)=(I_1^{*}(t)-I_2^{*}(t))/ (I_1^{*}(t)+I_2^{*}(t))$ is the fractional  deviation from the mean intensity, $\Delta \phi(t)=\phi_1(t)-\phi_2(t)$ is the relative phase delay, and $\alpha_{rot}$ the relative polarization rotation angle.\\

The null depth, defined as the inverse of the rejection ratio, is given by:
\begin{eqnarray}
N(t)=\frac{I_-^{*}(t)}{I_+^{*}(t)}=\frac{I_-(t)-I_{b}(t)}{I_+(t)-I_{b}(t)}
\label{nullgen}
\end{eqnarray}
where $I_{\pm}(t)$ are the constructive and destructive interference intensities including the background level and $I_{b}(t)$ is the background intensity collected by the interferometer. If $\Delta \phi(t)$, $\delta I(t)$ and $\alpha_{rot}$ are all $<<$ 1, the null depth for a point source in the absence of background can be approximated by:
\begin{eqnarray}
N(t) &\simeq& {1 \over 4} 
\left[ {(\delta I(t))^2 + (\Delta\phi (t))^2 + \alpha_{rot}(t)^2  } \right]
\label{null_func_0}
\end{eqnarray}

For a source of finite extent, the observed null depth also depends on the astrophysical null depth $N_a$, determined by the leakage of the source spatial brightness distribution through the null fringe pattern\footnote{For a given baseline orientation, the astrophysical null $N_a$ can be expressed in terms of the source complex visibility $\mathcal{V}$ as $N_a = (1-|\mathcal{V}|)/(1+|\mathcal{V}|).$}. For small values of $N_{a}$, the observed null depth can be expressed as \citep{Serabyn00} \footnote{The theory we present here can be extended for larger values of $N_{a}$, and of the error sources by keeping the full expression of $I_{\pm}^{*}$ in the definition of the null depth.}:
\begin{eqnarray}
N(t) &\simeq   \ N_{a} + {1\over 4}\left[ ({\delta I(t))^2} + ({\Delta\phi (t))^2} + {\alpha_{rot}(t)^2} \right]
\label{null_func_1}
\end{eqnarray}
Sometimes, interferometers do not measure the background intensity $I_{b}(t)$ nor the constructive interference term $I_{+}(t)$ at the same time as the destructive signal $I_{-}(t)$, but the observing procedure provides some estimates of their values which we denote as $\hat{I}_{b}(t)$ and $\hat{I}_{+}(t)$, while $\hat{I}_{+}^{*}(t)=\hat{I}_{+}(t)-\hat{I}_{b}(t)$. This means that one does not access the actual null, but an estimate of it given by:
\begin{eqnarray}
\hat{N}(t) = \frac{I_{-}(t)-\hat{I}_{b}(t)}{\hat{I}_{+}(t)-\hat{I}_{b}(t)} = N(t)\frac{I_{+}^{*}(t)}{\hat{I}_{+}^{*}(t)}+\frac{I_{b}(t)-\hat{I}_{b}(t)}{\hat{I}_{+}^{*}(t)}
\label{null_func_2}
\end{eqnarray}
or 
\begin{eqnarray}
\hat{N}(t) = I_{r}(t)N(t)+N_{b}(t)
\label{null_func_3}
\end{eqnarray}
where $N_{b}(t)=(I_{b}(t)-\hat{I}_{b}(t))/\hat{I}_{+}^{*}(t)$ is the background induced instantaneous error in the estimated null and where $I_{r}(t)=I_{+}^{*}(t)/\hat{I}_{+}^{*}(t)$ is the relative intensity deviation at time t.
\\
For small values of $N_{a}$, $\delta I(t)$, $\Delta\phi(t)$ and $\alpha_{\rm rot}(t)$, one can use Equation\,\ref{null_func_1} for $N(t)$, and the estimated null $\hat{N}(t)$ can be approximated by:
\begin{eqnarray}
\hat{N}(t) & \simeq & I_r(t)  \left[ N_{a} + {1\over 4}[({\delta I(t))^2} + ({\Delta\phi (t))^2} + ({\alpha_{rot}(t))^2}] \right]+N_{b}(t)
\label{null_func_4}
\end{eqnarray}
Although it does not correspond exactly to the actual instantaneous null level (which we cannot measure unless all peak and background measurements are made simultaneously), $\hat{N}(t)$ is the basic measured quantity derived from the observations which is used in  this paper.  All that matters for the accuracy of our data analysis is that we have: (i) the correct description of $\hat{N}(t)$ as a function of the astrophysical null  and instrumental noise terms, i.e.  Eq.\,\ref{null_func_4} and (ii) some way to evaluate these various noise terms  (or more exactly their distributions), which is the object of the following section.

\subsection{Analytical model for the statistical distribution of null values}
\label{Mathematical_considerations}

Because it would be extremely difficult to zero out or perfectly calibrate all instantaneous error terms, we take here the opposite tack, and ask what can be learned from the observed distribution of the null depth fluctuations. We thus begin by deriving the mathematical expression for the probability distribution corresponding to the null depth estimate given by equations \ref{null_func_2}, \ref{null_func_3} and \ref{null_func_4} when the relative phase, the intensity mismatch, the background, and the relative intensity all fluctuate randomly with small amplitudes. 

We first assume that the polarization term, $\alpha_{rot}(t)$,  is constant, so that we can neglect its time variability in the statistical analysis. For symmetrically placed beams within a common aperture, polarization mismatches should be small compared to phase and intensity errors, and this approximation is valid down to null levels of $10^{-4} $ or lower  \citep{Haguenauer06, Martin08}\footnote{For  long baseline interferometers, the polarization effect can be measured on calibrator stars and accurately corrected as it generally varies slowly over time.}. 
\begin{figure*}[!t]
\plotone{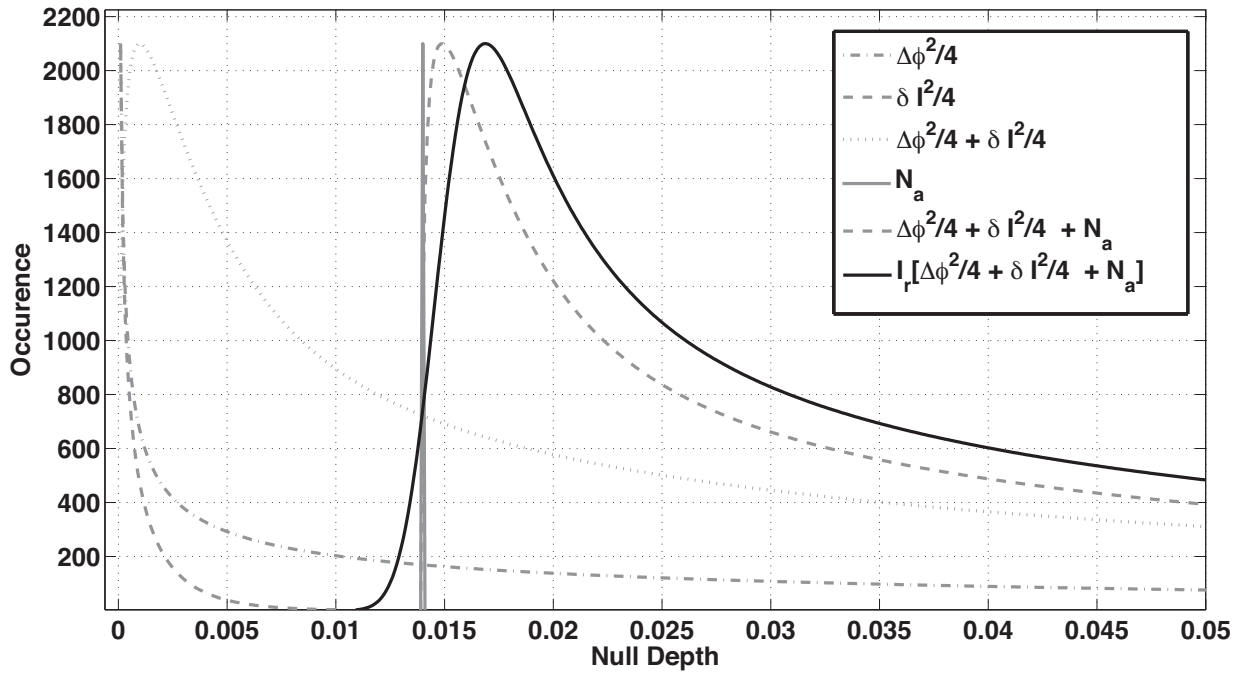}
\caption{Illustration of the construction of the null distribution from the individual contributions.  The phase and intensity terms (resp. $(\Delta\phi)^2/4$ and $(\delta I)^2/4$) are first summed which corresponds to the convolution of their respective probability distribution. The astrophysical null is then added. This step corresponds to the convolution by a dirac function. As a result, the probability distribution is translated horizontally by $N_a$. The last step consists in multiplying $(\Delta\phi)^2/4+(\delta I)^2/4+N_a$ by the relative intensity uncertainty $I_r$. The final distribution of the reconstructed null depth is represented by the black curve. The different curves represent realistic individual distributions of the phase and intensity errors, the astrophysical null and their sum. For each instrumental error terms $\Delta\phi(t)$, $\delta$I(t) and $I_{r}(t)$, a Gaussian distribution is assumed.}
\label{1}
\end{figure*}
Neglecting this term,  the measured null (Eq. \ref{null_func_4}) then consists of the sum of three terms multiplied by a fourth, and then the product is added to a fifth term.  Of these, only the astrophysical null term is fixed (for a given baseline vector). We next assume  that the remaining error terms in Eq. \ref{null_func_4}  - the relative intensity uncertainty $ I_r(t) $, the beam intensity mismatch $\delta I(t)$ , the beam differential phase $\Delta\phi(t)$ and the background uncertainty terms are uncorrelated random variables (this assumption is justified in Sect. \ref{section_correlation}). We further assume here that each of these  have normal distributions (see sections \ref{section_intensities}, \ref{section_background} and  \ref{section_opd}  for a complete description of the probability distributions of these terms), with means $\mu_i$ and standard deviations $\sigma_i$. Each individual probability density function (PDF) is then given by:
\begin{eqnarray}
f_{i}(z_i) = \frac{1}{\sqrt{2\pi}\sigma_i} e^{\frac{-(z_i-\mu_i)^2}{2\sigma_i^2}}
\label{pdf}
\end{eqnarray}
where the index $i$ refers equally to the $I_r(t)$, $\delta I(t)$, $\Delta\phi (t)$ and $N_b(t)$ distributions, and where $z_i$ is the corresponding random variable. 

However the $\delta I(t)$ and $\Delta\phi(t)$ distributions do not appear linearly but quadratically in the null distribution. In the case where $\delta I(t)$ and $\Delta\phi(t)$ both follow normal distributions, the PDFs of $(\delta I(t))^2 / 4$ and $(\Delta\phi(t))^2/4$ are given by:
\begin{eqnarray}
f_{i}(\frac{z_i^2}{4})=\frac{1}{\sqrt{2\pi} \sigma_i }\frac{e^{-(4 z_i+\mu_i^2)/2\sigma^2_i}} {\sqrt{4 z_i}} \cosh{\left(\frac{\mu_i\sqrt{4 z_i}}{\sigma^2_i}\right)}
\label{pdf_square}
\end{eqnarray}
\\
The two distributions, $(\delta I)^{2}/4$ and $(\delta\phi)^{2}/4$, are illustrated in Fig. \ref{1} for realistic values of their means and standard deviations. The next step in building the nulling PDF is to sum the phase and the intensity mismatch distributions. If $(\Delta\phi(t))^2$ and $(\delta I(t))^2$ are two independent random variables, the density function $f_{\delta I^2/4+\Delta\phi^2/4}(y)$ is given by the convolution of their respective density functions  \citep{Rohatgi76}. If we denote $y$ as being $(z_{\Delta\phi^2}^2+z_{\delta I^2}^2)/4$, this convolution can be expressed as follows:

\begin{eqnarray}
f_{\frac{\delta I^2}{4}+\frac{\Delta\phi^2}{4}}(y) & = & ( f_{\frac{\Delta\phi^2}{4}} \otimes   f_{\frac{\delta I^2}{4}} )(y)  \\
& = & \int_{-\infty}^{+\infty} f_{\frac{\Delta\phi^2}{4}}(y_1)f_{\frac{\delta I^2}{4}}(y-y_1) \, \rm d \it y_1
\label{pdf_sum}
\end{eqnarray} 


Adding the astrophysical null term, $N_{a}$, in Eq.\,\ref{pdf_sum} then corresponds to a further convolution of  Eq.\,\ref{pdf_sum} with a Dirac function $\delta(N_{a})$. The result is simply a translation of the density function by $N_{a}$ (see Fig\,\ref{1}): 
\begin{eqnarray}
f_{\frac{\delta I^2}{4}+\frac{\Delta\phi^2}{4}+N_{a}}(y)=f_{\frac{\delta I^2}{4}+\frac{\Delta\phi^2}{4}}(y-N_{a})
\label{pdf_na}
\end{eqnarray}
\\
Now folding the effect of the relative intensity uncertainty $I_r(t)$  into the expression for the measured null (Eq. \ref{null_func_2}), one computes the distribution of the product of $I_r(t)$ with $\left( (\delta I(t))^2/4+(\Delta\phi(t))^2/4+N_{a}\right)$. Assuming these are uncorrelated random variables \citep{Rohatgi76}, the resulting null depth distribution is: 
\begin{eqnarray}
f_{\hat{N}}(z_{I_r}) = \int_{-\infty}^{+\infty}\frac{1}{|y|} f_{\frac{\delta I^2}{4}+\frac{\Delta\phi^2}{4}+N_{a}}(y) f_{I_r}\left( \frac{z_{I_r}}{y} \right) \rm d\it y
\label{eq12}
\end{eqnarray}

The analytical solution for this integral exists for phase and intensity fluctuations following Gaussian distributions. However, the distribution $f_{\hat{N}}(z_{I_r})$ displays a singularity for $y=0$.



The final expression of the measured null distribution (see Eq. \ref{null_func_4}) is obtained by convolving Eq.\,\ref{eq12} with the equivalent background null depth distribution $f_{N_b}$:
\begin{eqnarray}
f_{\hat{N}}(N) = f_{\hat{N}}(z_{I_r})\otimes f_{N_b}(z_{N_b})
\label{pdf_final_bkg}
\end{eqnarray}

Summarizing all the steps described in this section, the final analytical expression for the measured null depth can be retrieved from the individual distributions as follows:
\begin{eqnarray}
f_{\hat{N}}(N) = f_{N_b}\otimes \left[ \int_{-\infty}^{+\infty}\frac{1}{|y-N_a|}f_{I_r}\left( f_{\frac{\delta I^2}{4}} \otimes f_{\frac{\Delta \phi^2}{4}} \otimes N_a \right)  \right]
\label{pdf_summary}
\end{eqnarray}
\\
The measured null distribution expressed by Eq. \ref{pdf_summary} depends on 9 independent parameters: the means and standard deviations of the 4 error distributions, and the astrophysical null.
In the simpler case of a system  where only a phase error impacts the measured null distribution, the null PDF depends only on three parameters: the mean and standard deviation of the phase error and the astrophysical null. Figure\,\ref{1} illustrates that $\mu_{\Delta\phi}$ and $\sigma_{\Delta\phi}$ define together the Full Width at Half Maximum (FWHM) and the skewness of the PDF, while $N_a$ only changes its horizontal position.

\subsection{Fitting strategies}
\label{section_fit}

Two methods can be used to generate null depth distributions to be fitted to the data. In the first one, referred to hereafter as the "analytical method", the distribution is generated analytically using the measured means and standard deviations of the background and intensity distributions. The second method, called the numerical  method, generates simulated distributions using the measured intensity and background distributions, together with  simulated phase error sequences having normal distributions, according to Eq.\,\ref{null_func_4}.  For illustration, we apply our techniques to data obtained with the Palomar Fiber Nuller (PFN), a deployable near infrared ($\simeq$ 2.16 microns) interferometric coronagraph developed at the Jet Propulsion Laboratory and recently installed at the Palomar Hale telescope \citep{Serabyn06a,Serabyn06b,Mennesson06,Martin08}. As described in the following sections, this method strongly reduces both statistical and systematic errors and can avoid the observation of calibrator stars (depending on the instrument). Therefore we call them respectively the analytical and numerical self-calibrated methods.

\subsubsection{Analytical Self Calibrated Method (ASC)}
\label{sect:anal_method}

The first strategy makes use of theoretical expression \ref{pdf_summary} to calculate null depth distributions, assuming that all instrumental terms follow gaussian statistics. No temporal data sequences are simulated, only the null probability distribution, which depends on nine parameters: the astrophysical null,  and the mean and standard deviations of the four gaussian error terms (relative phase, intensity mismatch, relative intensity and background variations). The number of  unknown parameters depends on the exact interferometric configuration and must be kept as small as possible in order to give a unique solution to the problem. In most interferometers, the individual beam intensities and the background intensity are monitored as part of the observing sequence, which leaves only three free parameters to be fitted: the mean and standard deviation of the phase error and the astrophysical null depth. In the case of the PFN, two symmetrically placed sub-apertures on the primary mirror are interfered, and using a rapidly spinning wheel, interleaved ($<$ 200 milli-sec) sequences of the interferometrically combined (nulled) signal, the individual beam intensities and the background are recorded.  Using this data, we fit the recorded relative intensity mismatch $\delta I(t)$, relative intensity $I_r(t)$ and background $N_{b}(t)$ with Gaussian profiles (see Sect.\,\ref{sect:limitations} for validation of this hypothesis). The resultant mean and standard deviation values are then injected into Eq.\,\ref{pdf_summary}. The remaining three free parameters of Eq.\,\ref{pdf_summary}, i.e. the differential phase parameters $\mu_{\Delta \phi}$ , $\sigma_{\Delta \phi}$, and the astrophysical null $N_a$, are then adjusted so as to fit the calculated curve to the observed null (Eq.\,\ref{null_func_2}) distribution. As detailed in section \ref{sect:unique},  except for the marginal case where the phase fluctuations are close to 0 (typically $\sigma_{\Delta\phi}<0.005 \,\rm rad$), only one combination of these 3 parameters provides a suitable fit to the observed null data distribution. The pair ($\mu_{\Delta \phi_d}$ , $\sigma_{\Delta \phi_d}$) defines both the FWHM and the skewness of the modeled null distribution, while $N_a$ adjusts the horizontal position of the distribution peak (see Fig.\,\ref{1}) and only one combination of $\mu_{\Delta\phi}$, $\sigma_{\Delta\phi}$ and $N_a$, fits the distribution. 
\\
Figure\,\ref{1} illustrates the construction of the analytical null depth distribution from the individual distributions while Fig.\,\ref{2} (left) illustrates the analytical fitting strategy on a nulling sequence measured on the sky with the PFN. After being fitted by Gaussian distributions, $\mu_{\delta I}$, $\sigma_{\delta I}$,  $\mu_{I_N}$, $\sigma_{I_N}$, $\mu_{N_B}$ and $\sigma_{N_B}$ are injected into Eqs. \ref{pdf} and \ref{pdf_square} to compute their impact  on the measured null distribution. The influence of the intensity mismatch is represented  by the grey long dashed curve in Fig.\,\ref{1}. Assuming a Gaussian distribution of the phase error as well, different values of $\mu_{\Delta\phi}$, $\sigma_{\Delta\phi}$ and $N_a$ are used to generate distribution curves. The impact of their distributions is also illustrated in Fig.\,\ref{1} by the grey dashed and dotted curve for the phase error and by the grey plain curve for the astrophysical null. All these distributions are finally combined together according to Eq.\,\ref{pdf_summary}. The resulting black curve can then be compared to the measured distribution (Fig.\,\ref{2}).

\subsubsection{Numerical Self Calibration Method (NSC)}
\label{section_num}

\begin{figure}[!t]
\includegraphics[height=7cm]{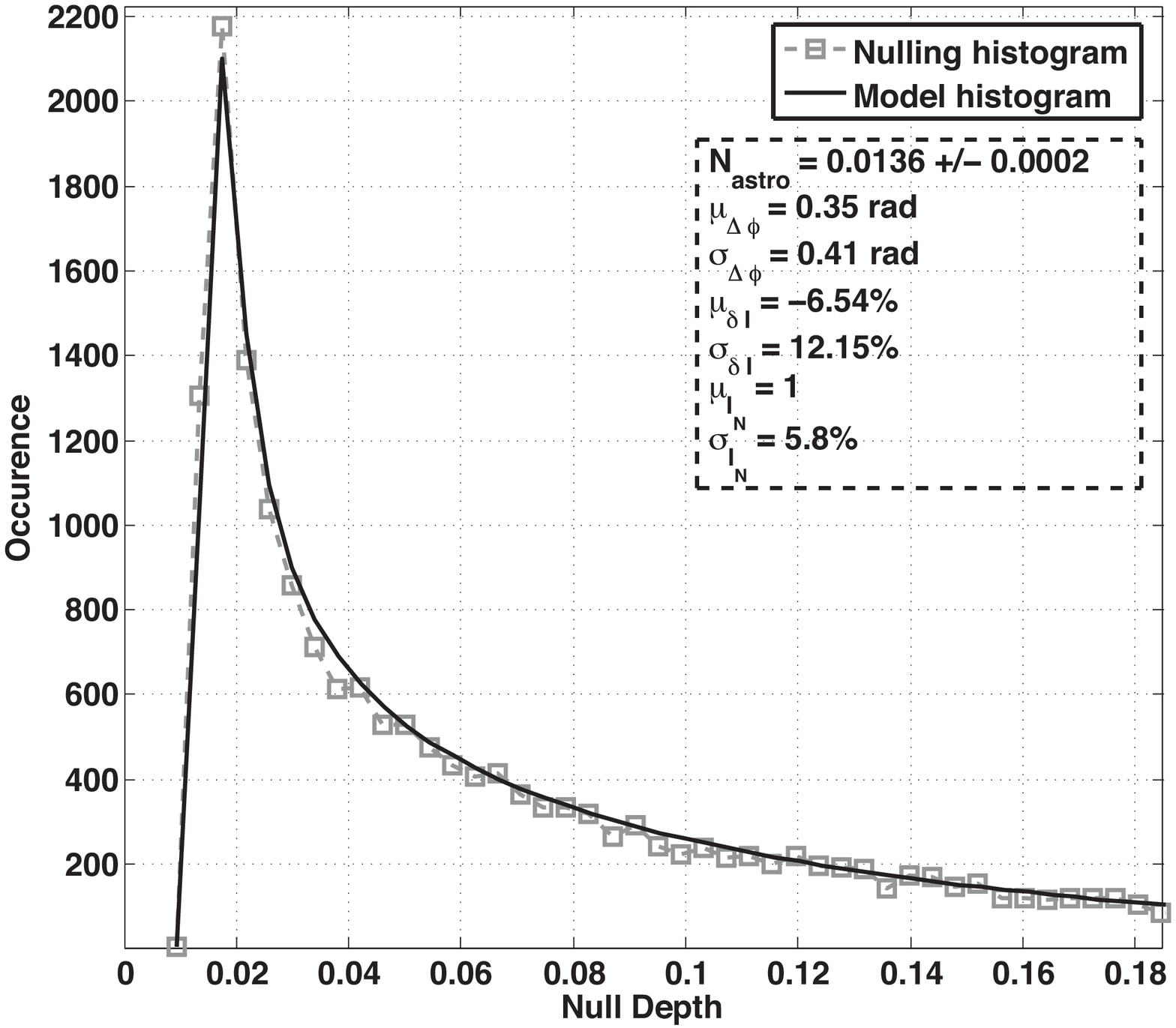}
\includegraphics[height=7cm]{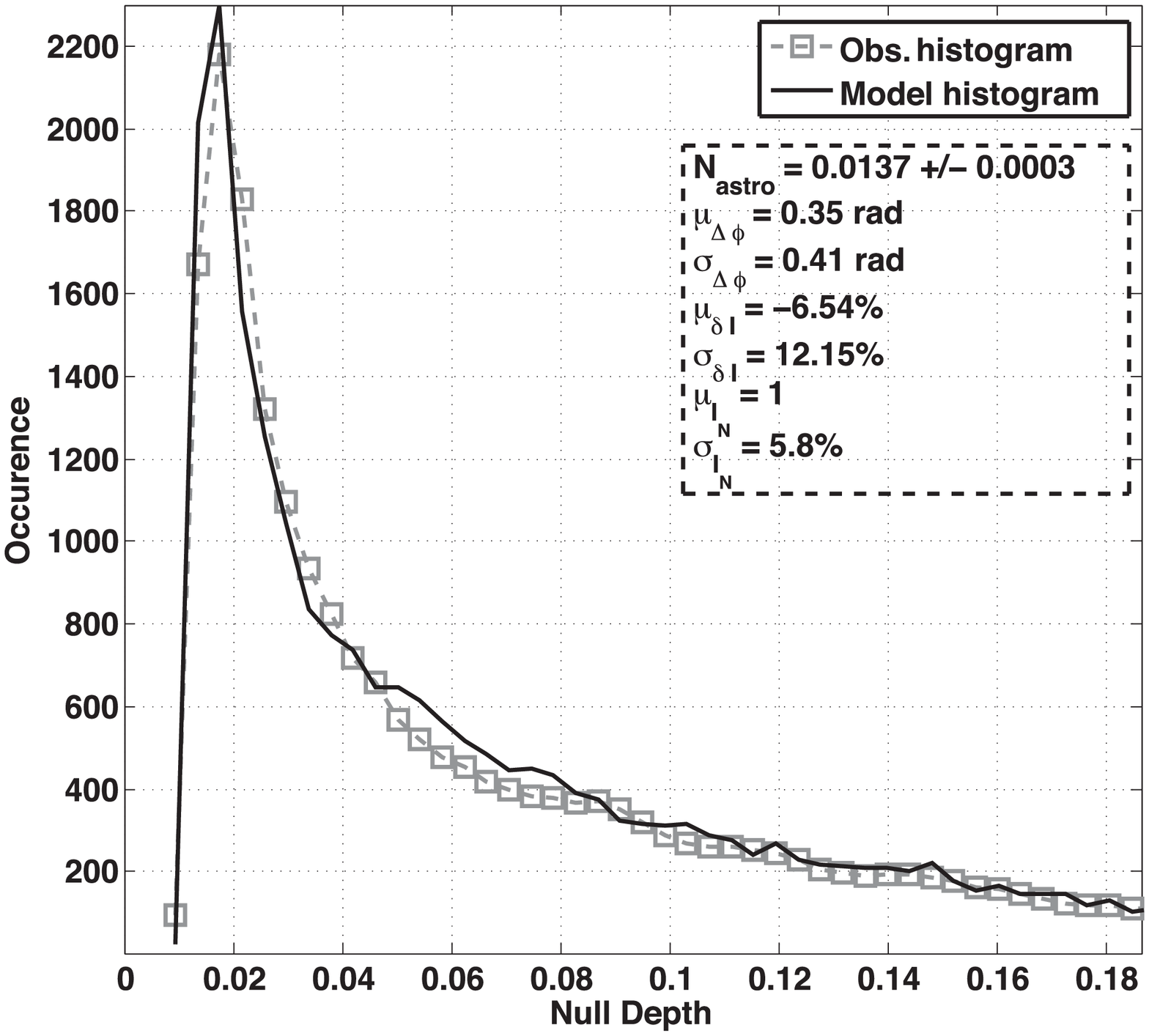}
\caption{\textit{Left}: Fit using the analytical self calibration method on a dataset obtained on $\alpha$ Boo with the PFN in July 2009. The astrophysical null corresponding to the best fit is $0.0136\pm0.0002$. \textit{Right}: Same fit but using the numerical self calibration approach. Note that the simulated distribution now presents more structure, as it integrates the actual distributions of background and intensity terms. The astrophysical null corresponding to the best fit is  $0.0137\pm0.0003$.}
\label{2}
\end{figure}

Unlike the ASC, the numerical self-calibration approach (NSC) does not make any assumption about the distributions of the intensity mismatch, background and total intensity terms, which are assumed to be measured within the null sequence or close in time. Instead of fitting the distribution of these three measured signals by gaussian distributions, we use the data - and hence actual distributions - recorded for each of these quantities and inject them directly into Eq.\,\ref{null_func_4}.

In the case of the PFN for instance, interleaved ($<$100\,milli-sec) measurements of the individual beams, interferometric (close to null) and background intensities are recorded over sequences of a few minutes. Although the background and individual beam signals are not recorded exactly at the same time as the null, their distributions can be measured with very high fidelity. In order to fit a sequence of observed null values, we combine these observed distributions with a generated random phase error (with a normal distribution)  of the same size (same number of data points) according to (Eq.\,\ref{null_func_4}). We only make two assumptions when using this method: (i) the differential phase follows a Gaussian distribution, (ii) the individual beam intensities are uncorrelated. The latter condition, which seems valid for the PFN measurements (see Sect.\,\ref{sect:limitations}), implies that the distribution of the differential intensity term $\delta I(t)$ derived from $I_1^*(t)$ and $I_2^*(t')$ measured at different times, is the same as if the individual intensities were measured simultaneously.  The main advantage of the numerical technique is that the data monitored by the instrument (i.e. the individual beam intensities and the background) are directly injected into the model. Therefore, no matter what the real distributions are for those terms, no bias is introduced into the modeled probability distribution. 
However, as the random generation of the differential phase vector produces slightly different distributions and best fit parameters for different seeds,  the numerical method adds some intrinsic uncertainty. This "fitting noise" is computed by generating many random phase errors and measuring the standard deviation of the resulting best fit astrophysical null depths. This uncertainty adds quadratically to the statistical error defined in the next section.  Consequently, the final error bar quoted on the astrophysical null derived by the numerical method is slightly larger than in the analytical case, but the potential sources of bias are reduced. Figure\,\ref{2}, right panel, shows an example of probability distribution fitting (same $\alpha$ Boo sequence as above) using the numerical  method.

\subsection{Error bars and residual comparison}

In this section, we compare the results obtained with the two fitting approaches. To perform this comparison, we make use of the retreived parameter corresponding to the best fits, the goodness of the fit and the relative residuals between the models and the data.  To compute the goodness of the fit and derive the optimum fit parameters, we minimize a reduced Pearson $\chi^2$ quantity, defined as:
\begin{eqnarray}
\chi^2 = \frac{1}{N_{\rm bins}-4} \sum_{i=1}^{n} \frac{\left( f_{\hat{N}}^{Obs}(i)- f_{\hat{N}}^{Theo}(i)\right)^2}{f_{\hat{N}}^{Theo}(i)}
\label{pearson}
\end{eqnarray}
where $f_{\hat{N}}^{Obs}$ and $f_{\hat{N}}^{Theo}$ are respectively the observed and theoretical null probability distributions and $N_{\rm bins}-4$ is the number of independent degrees of freedom. Following usual recommendations for robust fitting of probability distributions \citep{Cochran54, Rayson04}, we use a number of histogram bins equal to $\simeq\sqrt{N_{\rm pts}}$, where $N_{\rm pts}$ is the number of measurement points over the full range of observed null values. Also, only the largest null depth interval for which the occurence within each bin is $\ge5$ is used for the fitting. Unlike the NSC, the probability distribution obtained with the ASC method must be re-scaled prior to computing the $\chi^2$ to ensure that the total number of occurences in the theoretical distribution corresponds to the total number of measurements within the dataset. Mathematically, it comes down to introducing a scaling factor $C$ to match the integral of the observed and theoretical distributions over the domain of definition.  i.e. 
\begin{eqnarray}
C\, . \int_{N_{min}}^{1}f_{\hat{N}}^{Theo}(n) \, \rm d\it n = \int_{N_{min}}^{1} f_{\hat{N}}^{Obs}(n) \, \rm d\it n
\end{eqnarray}
where $N_{min}$ is the minimum observed null value of the distribution\footnote{The null depth in interferometry is generally considered to be defined between 0 and 1. However, the instantaneous measured null can be $<0$ because of the background fluctuations. This is why the limit of integration must be defined between the minimum measured null depth and 1.}. 

Overall, the analysis of different datasets with both fitting methods provided similar results, with reduced $\chi^2$ ranging between 1 and 1.5, meaning reasonably good statistical agreement between the model and the observations. The computation of realistic error bars must combine two different components which add quadratically: (i) statistical (random) errors and (ii) systematic errors.  Systematic errors, such as those arising from slow drifts in the experimental set-up (quasi-statics, e.g.  \citep{Colavita09}), are not captured by the statistical analysis of a single sequence, and will be discussed in Sect.\,\ref{section_comparison}. A thorough description of the different sources of quasi-static errors will also be presented in Sect.\,\ref{sect:limitations}. We only compute and quote statistical errors in this section. 

For an individual sequence, the statistical uncertainty $\sigma_{\rm stat}$ on the derived astrophysical null is assessed using the $\chi^{2}$ statistical properties (see \cite{num-recip}, \S 15.6.4). $N_{a}$ is varied around its optimal value while the $\chi^{2}$ is minimized by adjusting the other two parameters. The error bar on $N_a$ corresponds to the $N_a$ variation required to increase the reduced $\chi^2$ by a tabulated increment based on the desired confidence level and the number of degrees of freedom in the fit. A 68.3\% confidence level was adopted on the quoted error bar, and the  analysis of the covariance of the fit with the other two parameters (i.e. $\mu_{\Delta\phi}$ and $\sigma_{\Delta\phi}$) is presented in the Appendix. 
This estimation of the retrieved parameters error bars  is only valid if the observed null values are affected by zero mean gaussian noise. As a sanity check, we then also conducted a bootstrapping analysis -independent of the actual noise properties -, resampling and replacing the observed null values to generate many (500) "fake" sequences. Analyzing the corresponding histograms yields astrophysical null (68.3\% confidence interval) statistical uncertainties similar to those derived using the $\chi^{2}$ approach.

As an illustration, the left panel of Figure \ref{2} shows the best analytical self-calibrated fit (black curve)  to the null distribution observed (grey dashed line and squares) on  the bright star $\alpha$ Boo with the PFN (one particular two minute long sequence). The reduced $\chi^2$ is 1.17 and the derived astrophysical null calculated with a $1 \sigma$ confidence interval is  $0.0136\pm 0.0002$ (see Appendix \ref{app:a} for more details).The error bar quoted here is the statistical error only.  
The main advantage of this analytical fitting method is its mathematical consistency and precision. However, it assumes normal and uncorrelated noise distributions for all noise sources, instead of injecting their observed distributions into the model. These assumptions will be justified and explained in Section \ref{sect:limitations}. Another characteristic of this approach is that because of the $1/|y|$ term in Eq. \ref{eq12}, the distribution is not defined for a null depth N = 0. However this issue can be solved by simply removing the bin containing $N=0$ during the fitting process.  

\begin{figure}[!t]
\centering
\includegraphics[width=8cm]{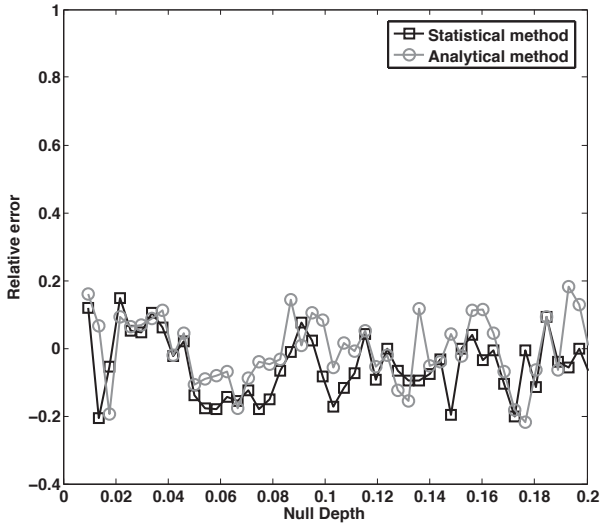}
\caption{Relative error ($(f^{Obs}_{\hat{N}}-f^{Theo}_{\hat{N}})/f^{Obs}_{\hat{N}}$) between the fitted null distributions and the measured one as a function of the null depth. The grey curve with circular markers, represents the relative error relative to the analytical self calibrated method while the black curve with square markers represents the relative error obtained with the statistical method. Both relative residuals are similar with rms values $\simeq$ 0.05. }
\label{rel_diff}
\end{figure}
The right panel of Figure\,\ref{2} represents the best fit obtained with the numerical approach on the same $\alpha$~Boo dataset. The reduced Pearson's $\chi^2$ is 1.23. The derived astrophysical null depth is $N_a=0.0137\pm0.0003$, in excellent agreement with the value obtained using the analytical approach and gaussian statistics for all instrumental terms. The quoted error bar accounts for both the statistical uncertainty and the numerical "fitting noise" discussed in section  Sect.\,\ref{section_num}

To complete this comparison, Figure \ref{rel_diff} shows  the relative difference between the measured null distribution and the distribution obtained with both the analytical  method (grey curve, circular markers) and the numerical fitting method (black curve, square markers). As can be seen on this figure, the two different statistical data reduction methods  are very equivalent in terms of accuracy: the relative residuals between their distributions and the measured one are similar. This is particularly true for small null depth values ($N<0.05$) where most of the astrophysical information is located. Overall, this comparison shows that very similar results are obtained on $\alpha$ Boo with the analytical an numerical methods.

\subsection{Amplitude of the fluctuations}

\label{sect:unique}

\begin{figure}[!t]
\epsscale{0.5}
\plotone{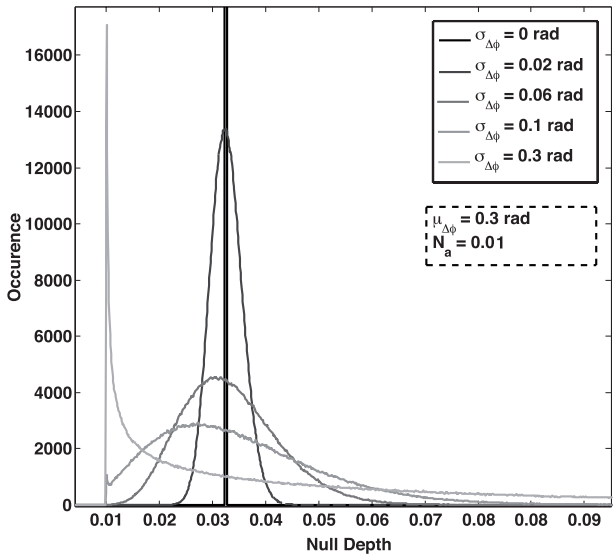}
\caption{Simulated null depth distributions for an astrophysical null of 0.01, a constant mean phase error of 0.3\,rad, and different values of the phase error rms. All the other error sources have been set to 0.}
\label{fig:phase_distr}
\end{figure}

\begin{figure}[!t]
\plotone{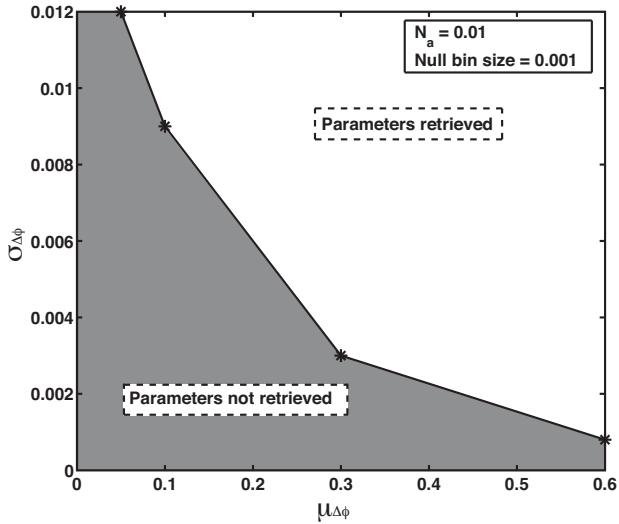}
\caption{For $N_a=0.01$ and a bin size of 0.001, this plot represents, as a function of the mean phase error, the minimum value of the phase rms required to fit unambiguously the distribution and retrieve the astrophysical null $N_{a}$. Note that on real data obtained with the PFN, we are located well within the \textit{parameters retrieved} zone (see Sect.\,\ref{sect:anal_method} and after). }
\label{7}
\end{figure}

 In this section, we demonstrate the conditions that must be fulfilled by the error fluctuations in order to produce a distribution that can be fitted by a unique combination of the parameters. For that, we consider the simpler case where only phase errors are present.
\\
First, let us consider the extreme case of a perfectly stable system ($\sigma_{\Delta\phi}=0$) but with an error on the phase shift ($\mu_{\Delta\phi}\neq 0$). The measured null distribution is then a Dirac function that peaks at $\hat{N}= N_a + (\mu_{\Delta\phi})^{2}/4$ (see Fig.\,\ref{fig:phase_distr}). Therefore, only the sum, $N_a + (\mu_{\Delta\phi})^{2}/4$, can be determined but not the astrophysical null.
\\
Of course, using a statistical approach for analyzing perfectly constant data does not make much sense and is not realistic. However, it shows that the phase fluctuations must have a minimal amplitude to make a statistical approach applicable. Now, let us consider the more realistic case of a system having both a phase fixed bias and phase fluctuations. Eq. \ref{pdf_square} expresses the impact of phase fluctuations on the null depth distribution. From this equation, it can be seen that the larger the fluctuations, the broader the corresponding distribution (see Fig.\,\ref{fig:phase_distr}). If the FWHM of this distribution is smaller than the bin size used for computing the null distribution, it appears as a Dirac function (which corresponds to a fixed phase error) and the fitted parameters cannot be found. Now, if the phase distribution can be properly sampled in several bins, the three parameters that must be fitted ($\mu_{\Delta\phi}$, $\sigma_{\Delta\phi}$ and $N_a$) can be retreived. More importantly, the solution found is unique. For small phase fluctuations, Eq. \ref{pdf_square} can be approximated by a Gaussian function whose FWHM is $2\sqrt{2 \ln 2} \times \sigma_{\Delta\phi}\sqrt{\mu_{\Delta\phi}}$. The criterion for a unique solution to our fit is therefore that this FWHM is larger that a few ($k$) times the histogram bin size (i.e. $k$\,bin~size $< 2\sqrt{2 \ln 2} \times \sigma_{\Delta\phi}\sqrt{\mu_{\Delta\phi}}$). From this equation, it can be seen that for larger mean phase offsets, the minimum phase fluctuation required to meet this criterion decreases. This is due to the fact that the null depth depends quadratically on the phase error. In practice, simulations have shown that  the phase distribution must be sampled over at least 6 bins ($k\ge 6$). Fig.\,\ref{7} shows, on simulated datasets,  the minimum amplitude of the phase fluctuations required as a function of the mean phase error for a bin size of 0.001. This bin size directly depends on the number of data points available within a dataset (see Sect. \ref{sect:anal_method}). It means that increasing the observing time (and therefore the number of data points within a dataset) allow the use of smaller bin sizes and hence even easier parameter retrieval. In this figure, a fit is considered successful when all three parameters are found within some a priori tolerable error. For $N_a$, it means that the error is smaller than the histogram bin size. For the two other parameters, it means that their effect on the null depth ($\Delta\phi^2/4$) is also smaller than the histogram bin size. It is interesting to note that even small fluctuations compared to the mean phase errors are sufficient to retrieve the astrophysical null depth with a very good accuracy. This also means that it is possible to measure $N_a$ even if the fluctuating phase error never reaches zero, and so even when the true astrophysical null value is never reached. Finally it is important to note that the results obtained with the PFN illustrating this paper correspond to parameter combinations located well within the \textit{"parameters retrieved"} zone of Fig.\,\ref{7}.

\section{On sky performance: classical vs. statistical reduction methods}
\label{section_comparison}

In order to investigate the validity and accuracy of our statistical  data reduction approach, we applied it to astronomical data obtained with the PFN during a July 2009 observing run. In order to evaluate the astrophysical null accuracy achieved with our statistical analysis, we present here the results obtained on a  series of consecutive independent measurements of $\alpha$~Boo with the PFN. We explore both the repeatability of the results (precision assessment), and their consistency with values previously reported by long baseline interferometry  (accuracy and bias assessment).

We use here a set of five independent null sequences recorded on $\alpha$ Boo with the PFN in July 2009, and compare the astrophysical nulls, $N_a$, and precisions derived from (i) the "classical" null (or visibility) data reduction method and (ii) from the probability distribution analysis. We then compare our results with the stellar diameter measurement obtained on this same star with long baseline interferometry (LBI), discussing the aspects of accuracy and systematic errors.

\subsection{Classical reduction method}

\begin{figure}[!t]
\plotone{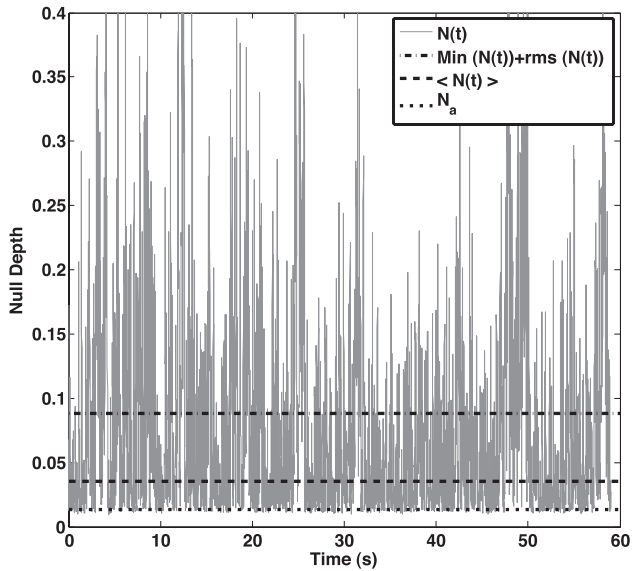}
\caption{Null depth fluctuations measured on the $\alpha$ Boo dataset. The dashed dotted line represents the highest null depth value which is taken into account in the classical data reduction approach. The dashed line corresponds to the mean null depth of the sequence prior to calibration ($\simeq0.035$) and the dotted line to the astrophysical null that is measured by the numerical statistical method ($0.0137\pm0.0003$). }
\label{fig:null_seq}
\end{figure}
\begin{figure*}[!t]
\epsscale{1}
\plottwo{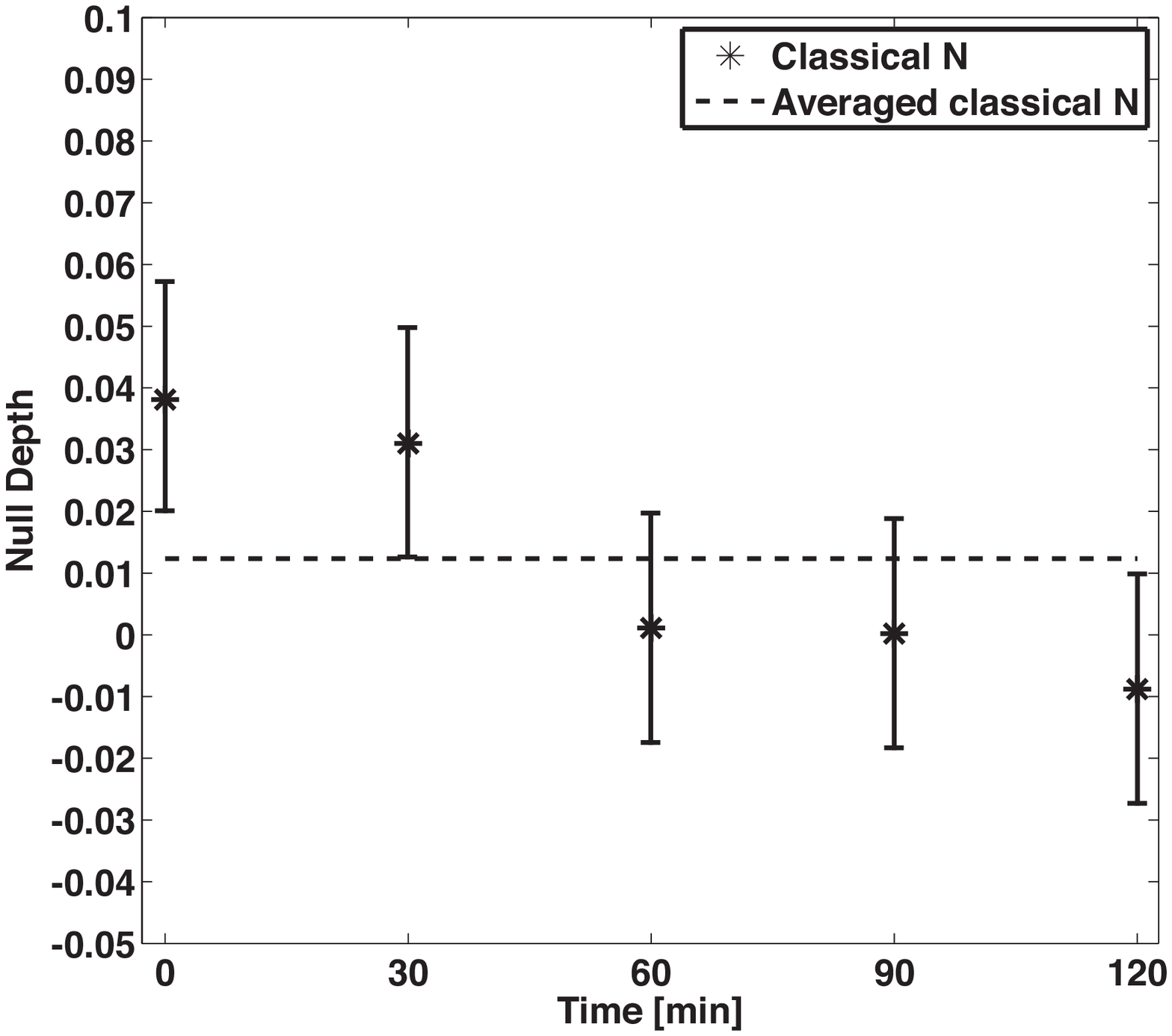}{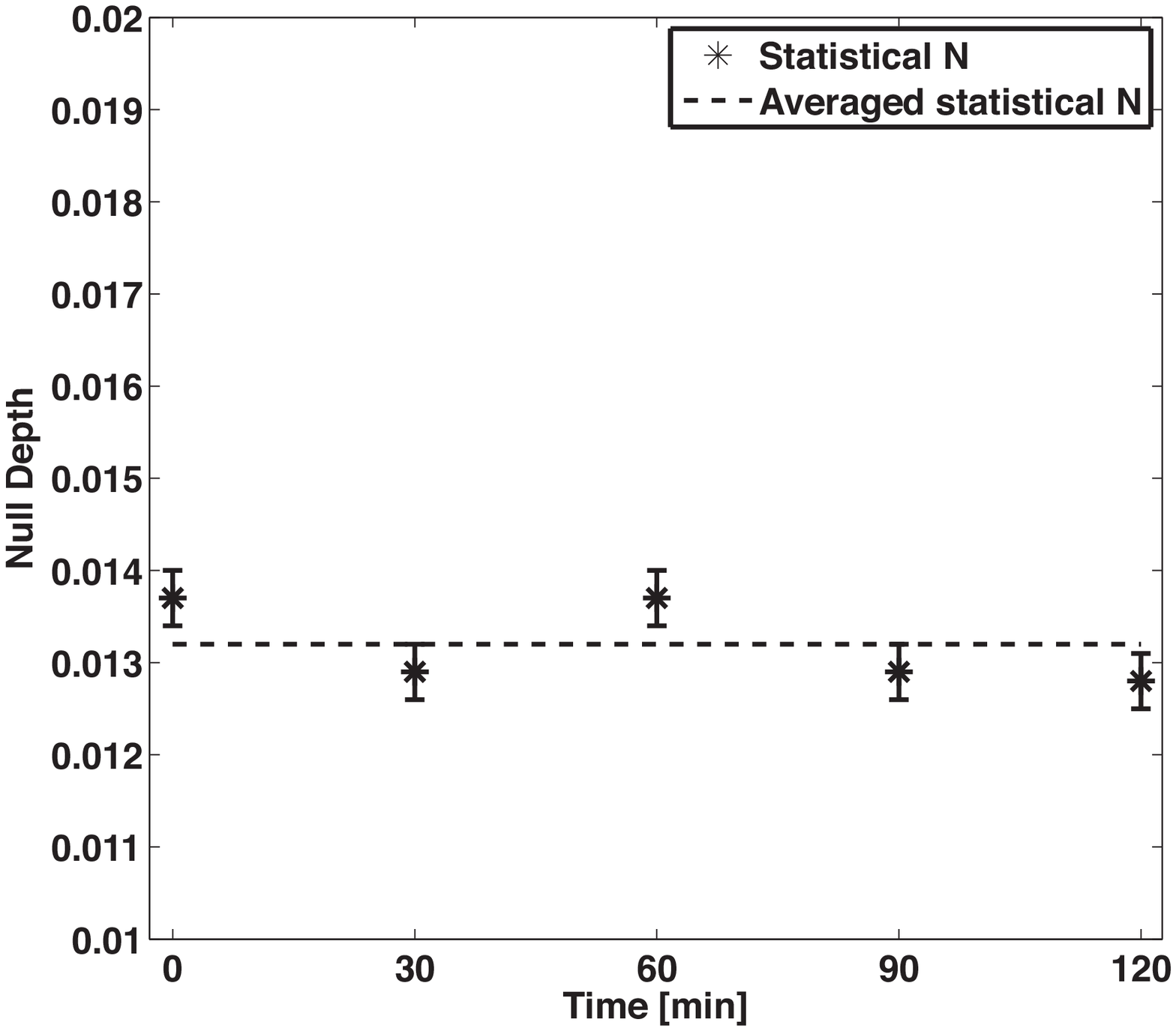}
\caption{Comparison between astrophysical null values obtained with both classical and statistical (numerical) data analysis approaches on $\rm \alpha \, Boo$ with the PFN. Left panel: results obtained using the classical reduction. The results drift significantly over time and the individual null depth error bars obtained on each dataset are 0.02. The mean measured astrophysical N is $0.0123\pm 0.008$. Right panel: results obtained using the statistical method. The measured nulls are very stable, with individual error bars around 0.0003. The mean astrophysical null measured is $0.0132\pm 0.00013$, where the error bar assumes no systematic uncertainties (see text for details). Note that the $y$ scale is different in the two figures.}
\label{5}
\end{figure*}

Very few nulling data from ground based telescopes have been analyzed so far, as only two nulling interferometers are operating: the Keck Interferometer Nuller \citep{Colavita09} and the BLINC Nuller \citep{Hinz00}. Until now, the method used to analyse nulling data was analogous to that used for calibrating visibility measurements. The principle is to first evaluate the null depth observed on the science target by averaging the fluctuating instantaneous null depth over a significant number of points. This measurement is  biased due to the fast fluctuations of phase and intensity errors. The same measurement is then conducted on a calibrator star of well known diameter, located close to the science target and with a similar magnitude at the wavelength of observation \citep{Merand05}. 
For both stars, the measured null depth  $\langle N(t) \rangle$ is the sum of the astrophysical null $N_{a}$ and the mean instrumental null $\langle N_{\rm i}(t) \rangle$ averaged over the sequence:
\begin{eqnarray}
\langle N(t) \rangle & = & N_{a}+ \langle N_{\rm i}(t) \rangle \\
\langle N_{\rm cal}(t  + \Delta t) \rangle & = & N_{a, \, \rm cal}+ \langle N_{\rm i, \,cal}(t + \Delta t) \rangle
\label{null_cal}
\end{eqnarray}
where the astrophysical null depth  on the calibrator star ($N_{a, \,cal}$) is assumed to be known thanks to an accurate photosphere model or from independent interferometric observations.  Therefore, assuming that the instrumental null is constant, one estimates the scientific target's astrophysical null as: 
\begin{eqnarray}
N_a = N_{a,\, \rm cal} + \langle N(t) \rangle -  \langle N_{\rm cal}(t+\Delta t) \rangle 
\label{null_sci}
\end{eqnarray}
Obviously, the accuracy on $N_a$ depends both on the calibrator's astrophysical null uncertainty and on the stability of the instrumental null (or the ability to extrapolate its value accurately based on bracketing calibrator observations).
\\
The method used to emulate a "classical analysis" of our null data  is the following. First the "bad" (large instantaneous nulls) data points within each dataset are rejected, both for the target and the calibrator. Only the data having null values between the minimum measured null $N_{min}$ and $N_{min}+\sigma_N$ are kept, where $\sigma_N$ is the rms of the null data (see Fig \ref{fig:null_seq}, dash-dot line). This is also called the sigma clipping method. The null depth of an individual object sequence is then computed as the mean of the remaining data points (see Fig \ref{fig:null_seq}, dashed line). The same approach is applied to both the scientific target and the calibrator data, and  the calibrated astrophysical null depth is then computed using Eq.\,\ref{null_sci}. The black stars in Figure\,\ref{5}, left panel,  represent the  calibrated null depths obtained with this classical data analysis on five consecutive $\rm \alpha \, Boo$ datasets. These data have been calibrated using five datasets obtained on $\alpha$ Her. The error bar on the individual  measurements is  given by the quadratic sum of the target statistical error, the systematic error  and the calibrator total error (statistical error and diameter uncertainty). The individual statistical error bars are computed from the variance of the null depth fluctuations within each dataset (after applying data clipping) and are equal to 0.018 on $\alpha$~Boo. The systematic error is more difficult to calculate and can be assessed both by comparing the measured null depth with the one expected from previous measurements of $\alpha$~Boo's stellar diameter and by comparing the individual statistical errors with the variance of the null over the 5 datasets. 
Assuming no/low systematics and averaging over the 5 data points, the astrophysical leakage measured on $\rm \alpha \, Boo$ is $0.0123\pm 0.008$  (see Table\,\ref{tab_summary}). The significant slow drift of the measured nulls in Fig.\,\ref{5}, left panel clearly illustrates that the classical method is very sensitive to the instrumental/ seeing conditions and to the fact that the calibrator was only observed {\it{after}} the five $\alpha$~Boo sequences and not in between them. The large error bars derived -even in the quite optimistic case of no systematics- demonstrate that in fact, with the short PFN interferometric baseline and when using the classical data reduction method, $\alpha$ Boo's near infrared diameter can not be measured reliably . 


\subsection{Statistical reduction method}

On the other hand, our statistical data analysis approach uses the whole range of null values recorded and neither uses nor requires any calibration star. Using the same five $\rm \alpha \, Boo$ datasets, the individual astrophysical nulls measured using statistics have much smaller individual error bars (0.0003), and are very stable over the whole two hours of observation (Fig \ref{5}, right panel). Using the 5 datasets obtained on $\alpha$~Boo, a set of $(N_{a,i}, \sigma_{\rm{stat},i})$ best fitting values is derived. From that ensemble, we compute the weighted mean value of $N_{a}$  with weights $w_i=1/\sigma_{\rm stat,\it i}^2$. The weighted mean astrophysical null value derived over the full sequence is 0.0132. Assuming {\it{no systematic errors}} and simply propagating the individual error bars ($\sigma_{\rm stat,\it i}$), the final statistical error bar is given by $\sigma_{\rm stat}^{-2}=\sum_{i}\sigma_{\rm stat,\it i}^{-2}$ and amounts to 0.00013. This yields an astrophysical null estimate of $N_{a}= 0.0132 \pm 0.00013$ for $\alpha$ Boo (see Table\,\ref{tab_summary}). 

Of course, systematic errors can be present in the data, for instance arising from slow drifts in the experimental set-up (quasi-statics) which are not captured by the statistical analysis of a single sequence. However, conversely to the classical method case, no obvious long term drift is visible versus time. The weighted standard deviation computed over the sequence is $0.0004$, in fairly good agreement with the quoted individual error of 0.0003, pointing to small systematics if any. This weighted standard deviation can also serve as an estimate of the systematic error per individual measurement, e.g. \citep{Colavita09}. The systematic error on the mean is likely smaller than that per individual measurement, but we do not have enough data to check for such reduction of the systematics wrt the number of measurements.  Consequently, we estimate the final error bar on $\alpha$ Boo's measured astrophysical null to be at the few $10^{-4}$ or lower.   

Another way to estimate systematics and constant biases is to compare the astrophysical null derived by the statistical method with previous measurements obtained by long baseline interferometry (LBI). This is the object of the following section. A detailed description of the potential sources of quasi-static errors as well as their impact on the null depth is also presented in section \ref{sect:limitations}. Finally, observations of calibrators can obviously be used in conjunction with the statistical data analysis to further reduce the effect of residual biases.

\subsection{Comparison to LBI data}

For a naked star represented by a limb darkened disk of diameter $\theta_{LD}$ with a limb darkening coefficient $A_\lambda$, the observed astrophysical null is given by \citep{Absil06a,Absil11}
\begin{eqnarray}
N_{a,LD} = \left( \frac{\pi B \theta_{LD}} {4\lambda}\right)^2 \left( 1-\frac{7A_\lambda}{15} \right) \left(  1-\frac{A_\lambda}{3} \right)^{-1}
\label{limb-darkened}
\end{eqnarray}
where $\lambda$ is the central wavelength of observation and $B$ the baseline length. For the PFN, these values are 2.16 $\mu m$ and 3.20\,m, respectively. This expression can be simplified in the case of uniform disk models by setting $A_\lambda=0$. 

LBI measurements of $\alpha$ Boo in the K band (where limb darkening effects and corresponding uncertainties are reduced) provide very accurate results. We use the value of 20.91 $\pm$ 0.08 \,mas derived by FLUOR/IOTA \citep{Perrin98,Lacour08}. This value is also very consistent with the previous measurement of 20.95 $\pm$ 0.20 \,mas obtained at I2T \citep{diBenedetto87}.

Using the 0.350 linear limb darkening coefficient predicted in the K band for a 4300K giant star with $\rm log\, g = 2.0$ \citep{Claret95}, we get an astrophysical null depth of 0.01314 $\pm$ 0.00010 at the PFN baseline. This is excellent agreement with or measured value of 0.001320 $\pm$ 0.00013 ( or $\pm$ 0.0004 when being conservative wrt systematics) reported above, which corresponds to a  limb darkened diameter of 20.96 $\pm$ 0.09 \,mas (see Table \ref{tab_summary}). The discrepancy between the PFN and LBI $\alpha$ Boo measurements is then at the $10^{-4}$ level, and within the error bars of each measurement.  This demonstrates that in the illustrative case of $\alpha$~Boo, our measurement is not only precise but also very accurate. It suggests that if any bias is present in our calibrator-free measurements of $\alpha$ Boo, they are at the few $10^{-4}$ level or below. A similar analysis of PFN data using the statistical reduction method confirms this result on a larger sample on 8 bright giants/ supergiants (Mennesson et al. in preparation). In comparison, the very best 1$\sigma$ null accuracy reported by long baseline interferometry is $\simeq$ 0.002 in the mid-infrared \citep{Colavita09}, and $\simeq$ 0.0025 in the near infrared \citep{Kervella04}, (equivalent to a visibility accuracy of 0.005 for an unresolved source). This indicates that using the self-calibrated data reduction approach, a gain of an order of magnitude in null (or visibility) accuracy can be achieved.


 In fact there is little that is specific to the PFN instrument in our approach, and the statistical data reduction method could in principle be applied to any 2-beam interferometer  working around null with a fringe tracker. Since null and visibility measurements are equivalent, the statistical analysis may thus also prove useful to regular long baseline visibility interferometry (Mennesson et al. in prep). 

\begin{table}[!t]
\caption{Comparison between  limb darkened (LD) diameters found by the PFN using both the classical and the numerical statistical data reduction method and by long baseline interferometry. Note that the null depth value given for LBI is an equivalent null on a 3.4\,m baseline derived from the measurement of the angular diamter. }
\begin{center}
\begin{tabular}{lcccccc}
\hline\hline
Method & Name & $N_a$ & $\theta [\rm mas]$  \\
\hline
Classic. nulls & $\rm \alpha\, Boo$ & $0.0123\pm 0.008$ & $20.25^{+6.4}_{-9.8}$ \\
Stat. nulls. & $\rm \alpha\, Boo$ & $0.0132\pm 0.00013 $ & $20.96\pm 0.09$   \\
LBI vis. & $\rm \alpha\, Boo$ & $0.0131\pm 0.00010 $ & $20.91\pm0.08$\\
\hline
\end{tabular}
\end{center}
\label{tab_summary}
\end{table}

\section{Possible limitations}
\label{sect:limitations}

We explore in this section some possible limitations of the statistical data reduction technique, which may appear when trying to measure very deep astrophysical null depths. Limitations arise from well identified sources: temporal effects, chromatic effects, and deviations from the assumptions used in the modeling. There are only two assumptions made in the self calibration technique: no temporal correlation between the individual beam intensities, and Gaussian distribution of the error sources \footnote{Note that in the case of the numerical method,  only the phase error distribution must be assumed to be Gaussian.}. In the following, we investigate these different effects, assess their contributions to the final null depth estimates, and suggest some mitigation techniques.   

\subsection{Intensity distributions}
\label{section_intensities}

\begin{figure*}[!t]
\plottwo{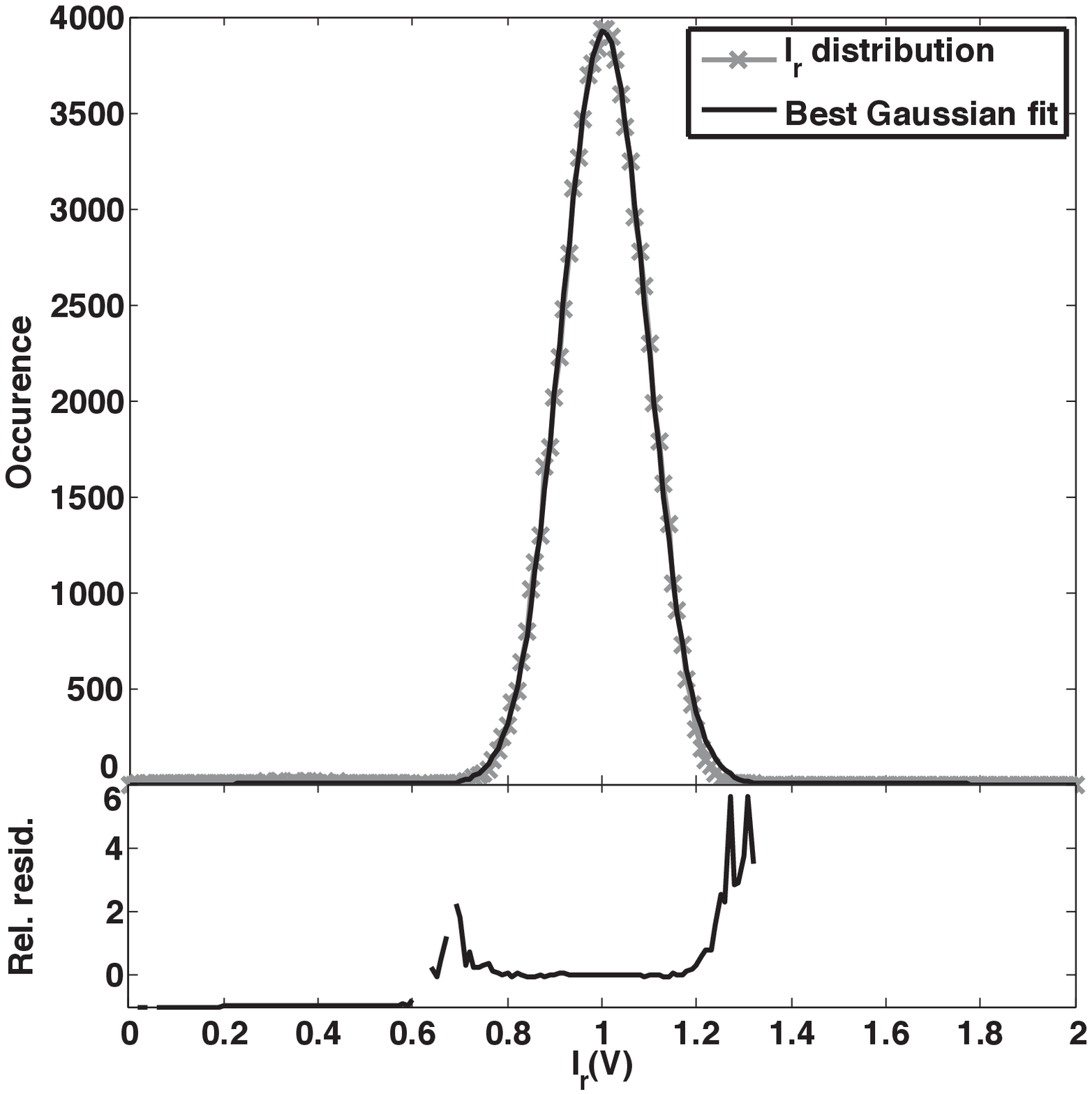}{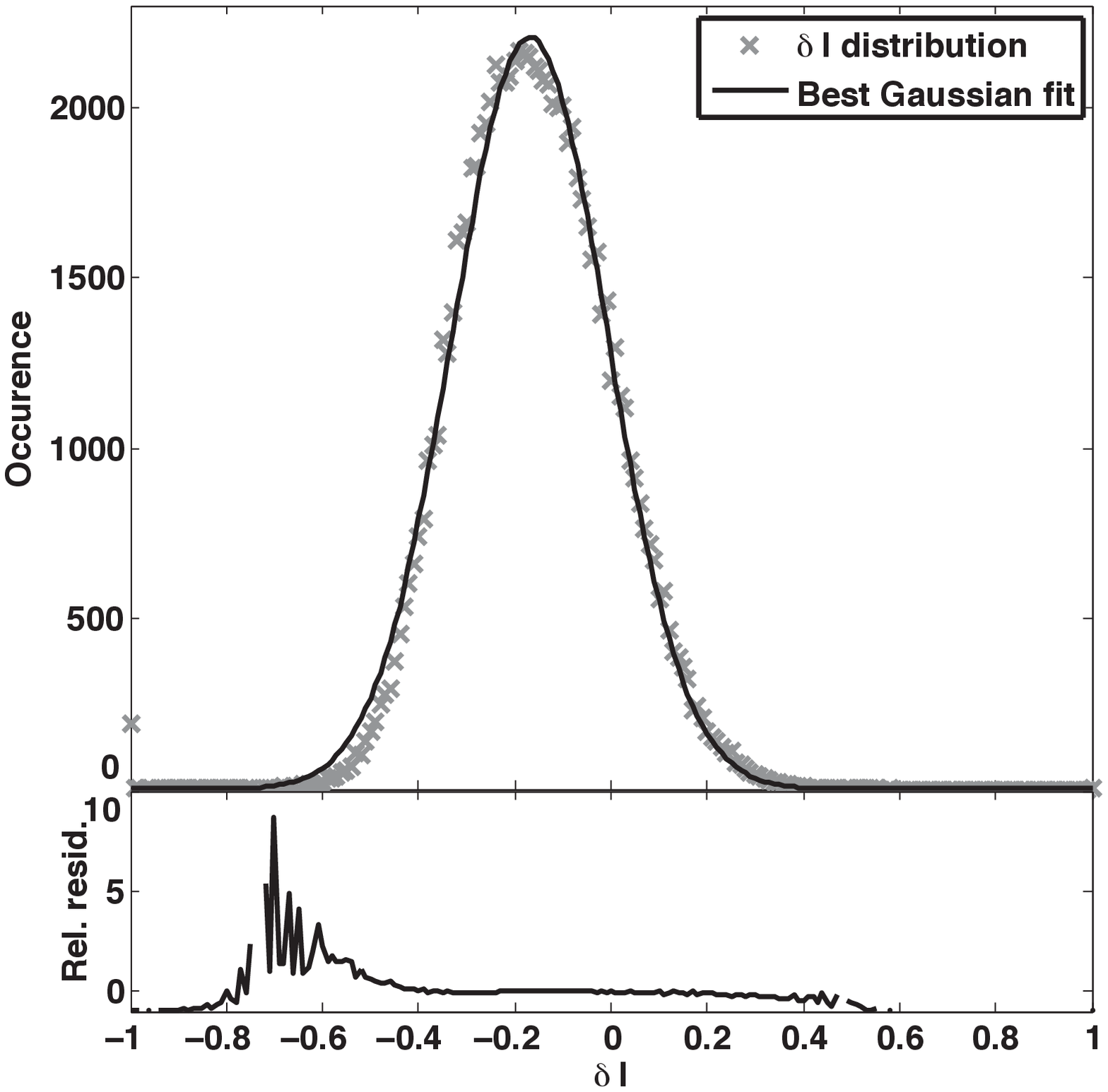}
\caption{Left: The top panel shows a comparison between the measured relative intensity uncertainty distribution (grey crosses) with the best fit of this distribution obtained with a gaussian distribution (black curve). The gaussian fit matches almost perfectly the measured distribution (except in the wings). The goodness of the fit is excellent with $\chi^2=0.9993$. Bottom panel: relative residual between the fit and the data ($(I^{Obs}-I^{Gauss})/I^{Obs}$).Right: The same comparison but for the relative intensity mismatch. The grey crosses represent the measured $\delta I(t)$ distribution while the black curve represent the gaussian distribution that best fits the measured distribution. Once again, the fit is excellent with a $\chi^2$ of $0.9952$. Bottom panel, relative residual between the observed intensity mismatch and the best Gaussian fit.}
\label{6}
\end{figure*}

Conversely to the numerical method, where the measured relative intensity uncertainty $I_r(t,\Delta t)$ and intensity mismatch $\delta I(t)$ are directly injected into the model, the analytical approach assumes these distributions to be Gaussian and computes their mean and standard deviation to feed the analytical expression of the measured null distribution (see Sect. \ref{Mathematical_considerations}). Therefore a possible limitation of the analytical approach could occur if these distributions are not Gaussian.  
\\
Fig.\,\ref{6} shows typical relative intensity uncertainty  and intensity mismatch distributions measured with the PFN. While the left hand panel compares the $I_r(t)$ measured distribution (grey crosses)  with a Gaussian distribution (black curve), the right hand one does the same for the $\delta I(t)$ distribution. As can be seen, both distributions can be reliably fitted by a Gaussian distribution. The goodness of the fit reduced $\chi^2$ values are $\simeq 0.99$ for both $I_r(t)$ and $\delta I(t)$. The bottom panel of these two figures illustrates the relative residuals between the observed distribution and the best Gaussian fit. For both fits, the residuals are close to zero for the whole central part where most of the information is located. Such a good agreement between the measured distributions and Gaussian distributions makes us confident these assumptions are justified and can be used but we note that a very slight skewness may be present.

\subsection{Background distribution}
\label{section_background}

\begin{figure*}[!t]
\centering
\includegraphics[width=9.5cm]{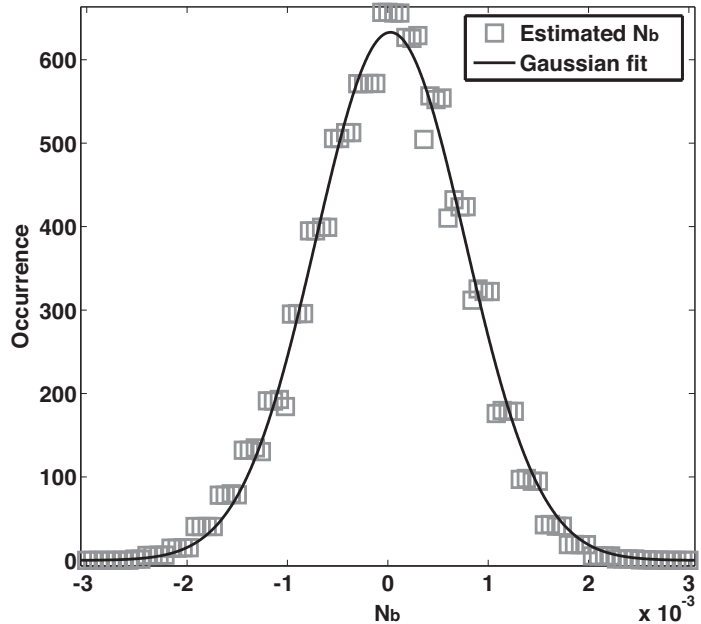}
\caption{Comparison between the distribution of the background induced instantaneous error and a Gaussian profile. The quality of the fit between the Gaussian model and the $N_b$ distribution is good with a $\chi^2=0.98$. }
\label{fig_background}
\end{figure*}

The analytical self-calibration method (unlike the NSC which uses the recorded background level), make the assumption that the distribution of the background level is normal and fits a gaussian profile on the recorded data to feed the analytical expression of the estimated null depth (Eq.\,\ref{pdf_summary}). However, background drifts can occur during observations either because of instrumental (e.g. electronic drifts) or observational reasons (the background depending on the sky position and time of observation) and can cause biases in the determination of the null depth. Figure\,\ref{fig_background} represent the distribution of the equivalent background null measured on $\alpha$~Boo over  2 minutes. The grey squares represent the actual measured distribution while the black curve is the best Gaussian fit corresponding to this distribution. Once again, the goodness of the fit is excellent with a $\chi^2\simeq 0.98$. However it must be stressed that this assumption is only verified for the particular PFN observations illustrated in this paper, and must be checked when using other instruments.

\subsection{Correlation issues}
\label{section_correlation}

In our statistical (both numerical and analytical) self-calibrated method (section 2.2.1), we made the assumptions that the different noise terms (background, differential intensity, overall intensity and differential phase) were temporally uncorrelated, so we could compute the theoretical null distribution from the individual noise distributions. 
\\
The cross-correlation of the intensity and phase terms is difficult to estimate. However, the optical/ near infrared coherence length of the atmosphere is generally much smaller than the distance between an interferometers sub-apertures.  Consequently, as the interferometric baseline increases, an even smaller correlation is expected between differential phase and intensity. Even with the compact PFN system, the typical value for the Fried's radius is 70 cm \citep{Roddier83} at 2.2 microns, to be compared with an interferometric baseline of 3.4 m. In the case of single-mode fiber injection,  the intensities of the individual beams are primarily driven by the local tip-tilt and overall phase corrugations of the individual apertures, and have no relation to the differential phase between the apertures. This suggests that the absence of correlation between the different noise terms is to first order justified both for the PFN, and long baseline interferometry in general.
\\
\begin{figure}[!t]
\begin{centering}
\includegraphics[width=10cm]{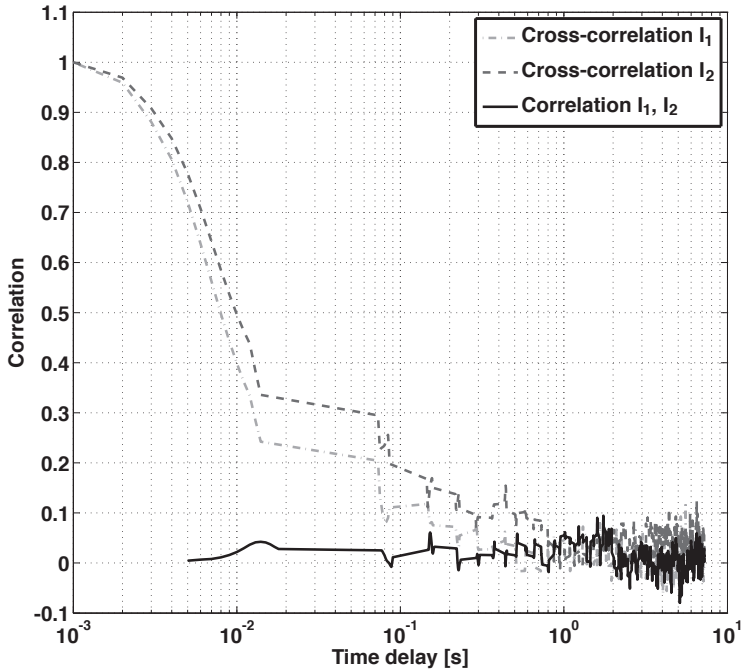}
\caption{Typical intensity correlation measured during an observation with the PFN. The dashed dark-grey line correspond to the correlation between the beam1 intensity at time $t$ ($I_1(t)$) and the same beam intensity at time $t+\Delta t$ ($I_1(t+\Delta t)$). The dashed light-grey line represent the same correlation but computed for beam 2 and the black line is the correlation between the two different beams intensities for different time delays.}
\label{fig:int_correlation}
\end{centering}
\end{figure}

The actual amount of correlation between the two beam intensities can be assessed by comparing the correlation of $I_1(t)$ with $I_1(t+\Delta t)$, $I_2(t)$ with $I_2(t+\Delta t)$ and $I_1(t)$ with $I_2(t+\Delta t)$. Figure \ref{fig:int_correlation} illustrates such a comparison for a typical dataset obtained with the PFN. For time delays close to zero both beam intensities are of course perfectly correlated with themselves (dark and light grey dashed curves). The correlation then decreases following a gaussian like curve until typical time delays of $\sim0.2$\,s are reached. The correlation is then very close to  zero ($<$ to a couple of percent). This information directly gives us an indication of the atmospherical conditions. Indeed, as long as the turbulent cells stay above the individual apertures, some correlation will remain between the beam intensity measurements at times $t$ and $t+\Delta t$. Considering that our apertures are 1.5\,m wide, we expect to lose completely the correlation between $I_1(t)$ (resp. $I_2(t)$) and $I_1(t+\Delta t)$ (resp. $I_2(t+\Delta t)$) when $\Delta t$ is such that the turbulent cell has moved by more than $1.5$\,m. Given the correlation time obtained from Fig.\,\ref{fig:int_correlation}, we can infer a wind speed during the observations of $\sim7.5$\,m/s which is consistent with typical conditions at Palomar Observatory. On the other hand,  the profile of the correlation between $I_1(t)$ and $I_2(t+\Delta t)$ is completely different. Indeed, the measured values are always under $5\%$, even at short time delays. We can therefore quantitively confirm that even for interferometric observations with small baselines and operated under good atmospheric conditions, no significant correlation exists between the two beam intensities. 

There is no physical reason why the background should correlate with any of the other terms. However it is possible that background intensity and the beam intensities are correlated to some extent if they are measured sequentially on the same detector (remanence). Such an effect depends on the hardware used for each instrument. We have computed its effect on the PFN measurements by computing the correlation between the mean beam intensities and background measurements over each chop cycle. We find that the correlation, if any, is smaller than $5\%$.

\subsection{Differential phase distribution}
\label{section_opd}

For both statistical reduction techniques presented,  the differential phase - computed at the central observing wavelength, see section \ref{sect:chrom}- is assumed to exhibit a Gaussian distribution over the recorded nulling sequence. The validity of this assumption is difficult to assess from our data. As long as the instrument tracks around a constant optical path difference (OPD) position, it seems a reasonable assumption. In the case of the PFN, the two beams come from the same AO corrected wavefront. Tracking a single OPD comes down to the fact that the AO system, which essentially acts as a fringe tracker, tries to maintain the same reference flat wavefront over the sequence. Some studies have shown that indeed, the phase residuals after a AO system are Gaussian, which supports our assumption \citep{Cagigal00}. If for some reason the fringe tracker or AO system loses lock, or if the OPD is obviously oscillating between several distinct positions, the resulting distribution will no longer be Gaussian, and the corresponding data should be discarded. The reduced Pearson $\chi^{2}$ defining the quality of the probability distribution fit (Eq. \ref{pearson}) is a good quantitative tool to assess the validity of the gaussian OPD distribution. If the measured $\chi^2$ are much larger than one, the error bars on the final astrophysical ND should be increased accordingly. Determining the potential bias caused by any departure from a Gaussian OPD distribution is beyond the scope of this paper, but can likely play a role for measuring reliable nulls at very low levels.

\subsection{Temporal effects}

The nulling expression established above (Eq. 8) is valid for instantaneous nulls. However, a photometer or camera will work with a limited frequency response or a finite individual integration time $\delta t$. In practice, this means that even when all of the instrumental terms of Eq 4 go through zero instantaneously, the measured null will in general be higher. Assuming that the polarization mismatch term is negligible, the best measurable null at any time t will be limited to:

\begin{equation}
N_{min}= \frac{ \sigma^2_{\delta I(t, dt)} + \sigma^2_{\Delta \phi(t, dt)} } {4}
\end{equation} 
\\
where $\sigma^2_{\delta I(t, dt)} $ and $\sigma^2_{\Delta \phi(t, dt)}$ are respectiveley the variance of the intensity mismatch and of the differential phase, both measured over a time interval $\delta t$. The effect of finite temporal integration is then to cause a (positive) bias to the observed null depth.  If the individual integrations are short enough compared to the typical fluctuation timescale, this bias can be kept to a very low level. Moreover, it could be at least partially calibrated via observations of reference stars. In the case of the PFN for instance, we use 2-10 ms individual integrations, to be compared with $\simeq$ 100 ms for the typical coherence time of atmospheric turbulence at K band. Using a Kolmogorov spectrum for the turbulence, and using the PFN short baseline, we find for instance that the atmospheric phase rms is less than 1 nm over 10 ms, limiting the minimum null depth   $\simeq 2 \times 10^{-6}$ . Similarly, the intensity mismatch term follows atmospheric timescales, and its variance over 10 ms is not expected to cause any significant bias either.  Laboratory nulling experiments with fiber nuller set-ups have already produced 10 ms nulls at the $\simeq 10^{-6}$ level with visible laser light \citep{Haguenauer06}, and $10^{-4}$ nulls with dual polarization broad-band light over the full  K band. In the latter case, dispersive and/ or polarization effects are likely dominating the error budget, and the effect of finite integration is not found to play a role up to 50 ms. Finally, astrophysical nulls at the 0.001 (or even slightly lower) level have been measured on Vega with the PFN (Mennesson et al. , in prep.), showing experimentally that temporal effects are at most at this  level (and probably much smaller) on the PFN. The optimum individual integration time is thus a trade-off between sensitivity and  dynamic range.

\subsection{Chromatic effects}
\label{sect:chrom}

Usually, interferometric / nulling observations are conducted over a finite spectral bandwidth. We concentrate here on the effects of the chromatic phase term, expected to dominate over the chromatic  aspects of intensity or polarization mismatch.
For a polychromatic observation, the phase error ($\Delta \phi(t)$) is the sum of the piston error calculated at band center ($\Delta\phi_c(t)$) and the chromatic phase error ($\Delta\phi_\lambda(\lambda,t)$). \citet{Serabyn00} has demonstrated that the influence of these phase errors on a polychromatic null depth measurement is given by

\begin{eqnarray}
N_\phi(t) = \frac{\Delta \phi_c^2(t)}{4} +  N_{\rm chrom}
\label{eq_phase_error3}
\end{eqnarray}
\\
where $N_{chrom}$  = $\int_{\lambda_{min}}^{\lambda_{max}}\frac{\Delta\phi_\lambda^2(\lambda, t)}{4}\, d\lambda$ is the chromatic null bias . So even in the case where the differential phase at the center of the band is zero, a positive bias is present (either constant or slowly varying, see below) , and one measures $N_\phi(t)= N_{chrom}$. This additive bias directly impacts the astrophysical null depth measurement.

In the case of the PFN, this chromatic term is minimized by inserting  glass plates of different thickness in each of the two beams. The chromatic bias is experimentally found to be lower than $10^{-4}$ in the laboratory.  On the sky, the dispersive phase  is no longer a strictly static term coming from the instrument. It is also impacted by differential atmospheric refraction across the band, and varies over the night according to the target's position with respect to zenith.  Detailed calculations are beyond the scope of this paper, but this effect is small ($<10^{-4}$) across the K band with the PFN short baseline when observing within 20 degrees of zenith.  Additionally, solutions exist to strongly reduce or completely eliminate this effect: disperse over several spectral bins, always orient the interferometric baseline perpendicular to the refraction direction (trivial on a single-dish interferometer with multiple sub-apertures), or use atmospheric dispersion compensators at the telescope.  Moreover, this refraction effect is fortunately very repeatable, and can be precisely calibrated by observing reference stars at the same zenith angle.

\subsection{Summary of limitations}

The assumptions proper to the analytical method (gaussian distribution of background and intensity terms, correlation issues) seem all individually valid in the case of the PFN. The analytical method also provides very similar results to those obtained by the numerical method, which makes fewer assumptions. The assumption that the differential phase follows a gaussian distribution can not be directly checked with the PFN data, but seems reasonable wrt theoretical expectations. 

Temporal and chromatic effects (as well as polarization effects, which we completely ignored for the PFN), may come into play at the $10^{-4}$ level, even more when considering the application to LBI which uses very long non common beam paths. However, these systematic effects - slowly varying for the most part-, can be either minimized by instrumental design, or strongly reduced via observations of calibrator stars.  

A more serious limitation to the reduction method presented is that its ultimate sensitivity may be limited by the small integration times needed to freeze the phase and intensity fluctuations. Infrared cameras with very low read noise will definitely help. Taking long sequences will also help, up to the point where systematics will dominate. More work is clearly needed to understand the trade-off between individual integration time, sensitivity and final accuracy.





\section{Conclusions}

The theory of a new data reduction method for interferometric nulling (or visibility) observations  has been presented in this paper. Based on the analysis of null distributions, this technique allows the retrieval of high dynamic range astrophysical null depth measurements, at contrast levels far exceeding the usual limits set by mean instrumental performance and fluctuations.  The ultimate performance of this statistical data reduction depends on the specific design of the interferometric instrument and on the observing strategy. This technique is potentially applicable to any interferometric set-up using a fringe tracking capability and any type of beam recombination (co-axial or multi-axial) into a single-mode waveguide. 
Applying our data reduction method to stellar observations obtained at K-band ($\simeq$ 2.2 microns) with the first generation fiber nulling instrument installed at the Palomar 5\,m (Hale) telescope, we demonstrated for the first time that: (i) deep and accurate nulling is not restricted to mid-infrared wavelengths but may be extended to the near infrared domain, providing substantial gains in resolution and sensitivity and (ii) nulling accuracies significantly lower than $10^{-3}$ (systematics and statistical errors included) can be achieved, without any observation of calibrator stars. Although this remains to be further validated with an optimized instrument,  simulations suggest that this new analysis will enable direct detection of faint structures at the $\simeq10^{-4}$ level within the near diffraction limit of large AO equipped ground based telescopes, i.e at angular separations ranging from $\simeq$ 20 to 150 mas. Implications for  high accuracy long baseline interferometry, both from the ground and from space, remain to be quantitatively explored. But since the statistical approach allows the detection of astrophysical signals well below the mean contrast level {\it{and its rms fluctuations}},  we anticipate that the instrumental stability requirements could be strongly relaxed. This implies that the constraints on intensity and phase fluctuations may be strongly reduced. This is a most attractive prospect for deep nulling interferometry from space. A similar statistical analysis may also be conducted successfully for regular coronagraphic instruments \citep{Riaud10}.


\acknowledgements{
This work was carried out at the Jet propulsion Laboratory, California Institute of Technology, under contract with the National Aeronautics and Space Administration. The data presented in this paper are based on observations
obtained at the Hale Telescope, Palomar Observatory, as part of a continuing collaboration between Caltech, NASA/JPL and Cornell University.  We wish to thank the Palomar Observatory staff for their assistance in mounting the PFN and conducting the observations at the Hale telescope. The research was supported by the Fond National de la Recherche scientifique de Belgique (FNRS), by the  Fonds pour la formation \`a la Recherche dans l'Industrie et dans l'Agriculture de Belgique (FRIA) and by the Communaut\'e Francaise de Belgique - Action de recherche concert\'ee - Acad\'emie Wallonie - Europe and by the Center for Exoplanet Science. The authors thank the referee for a careful review and for giving us relevant comments that helped to significantly improve the paper.
} 

\appendix
\section{Confidence intervals and covariance}
\label{app:a}

\begin{figure}[!h]
\includegraphics[height=4.5cm]{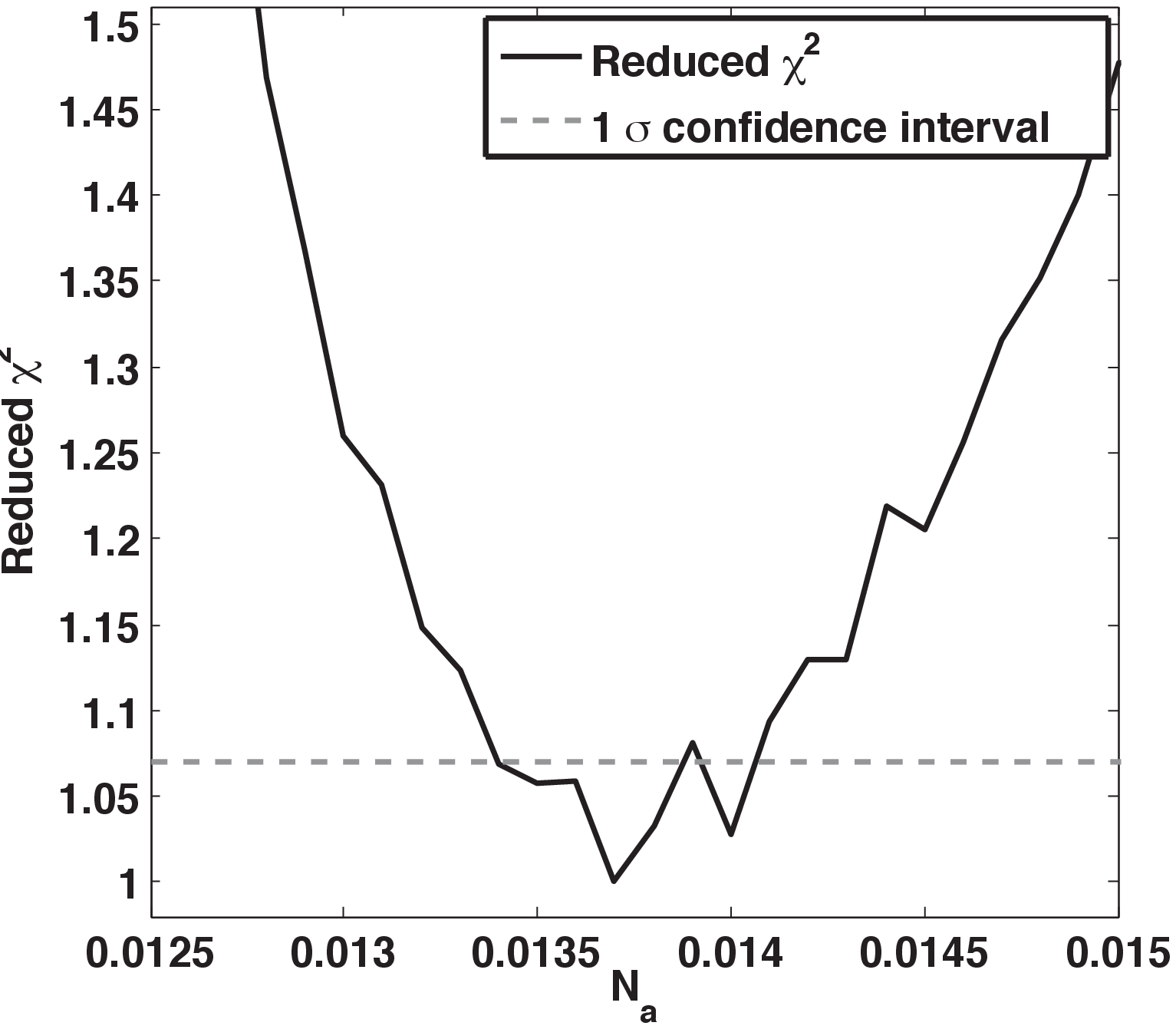} 
\includegraphics[height=4.5cm]{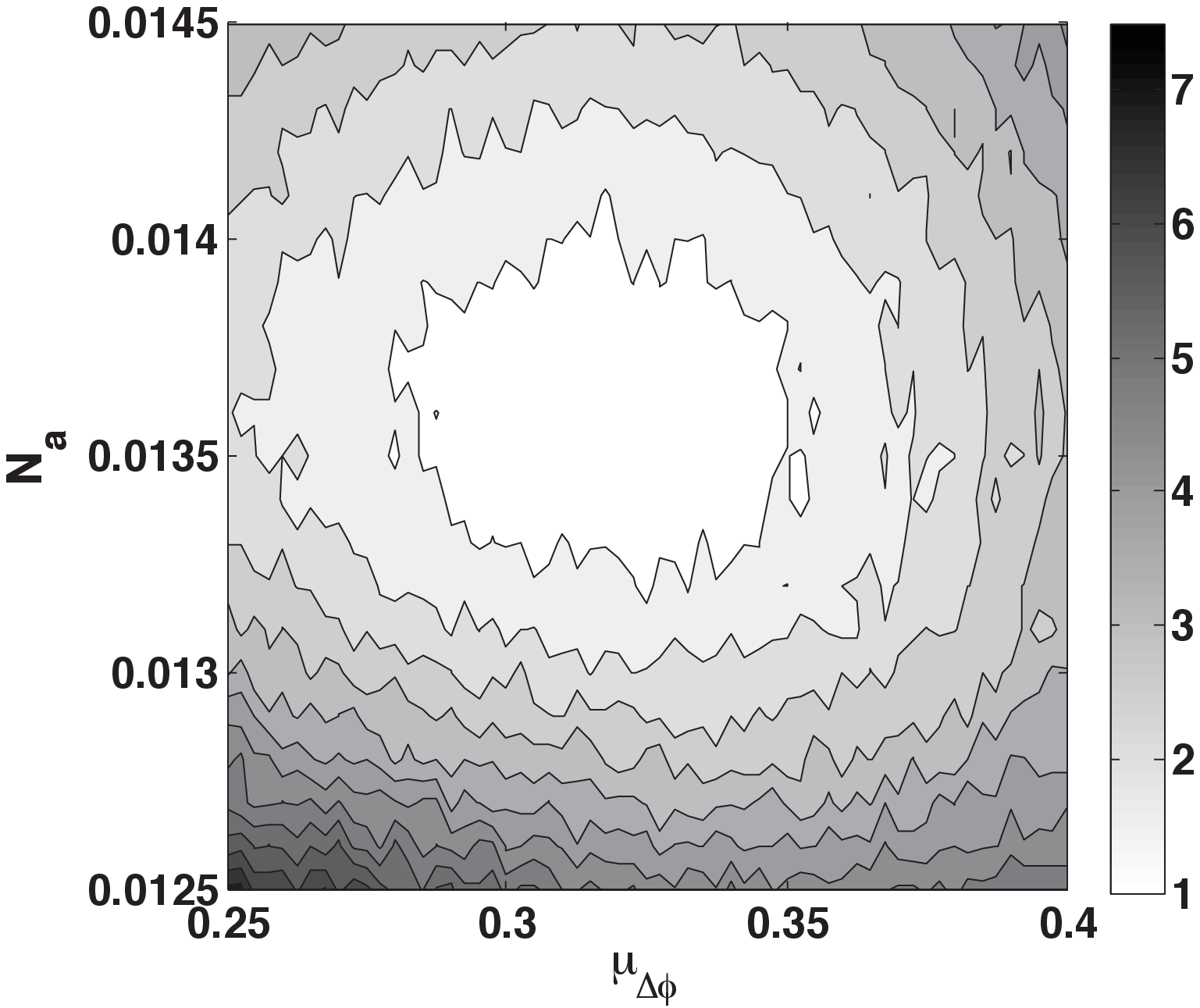} 
\includegraphics[height=4.5cm]{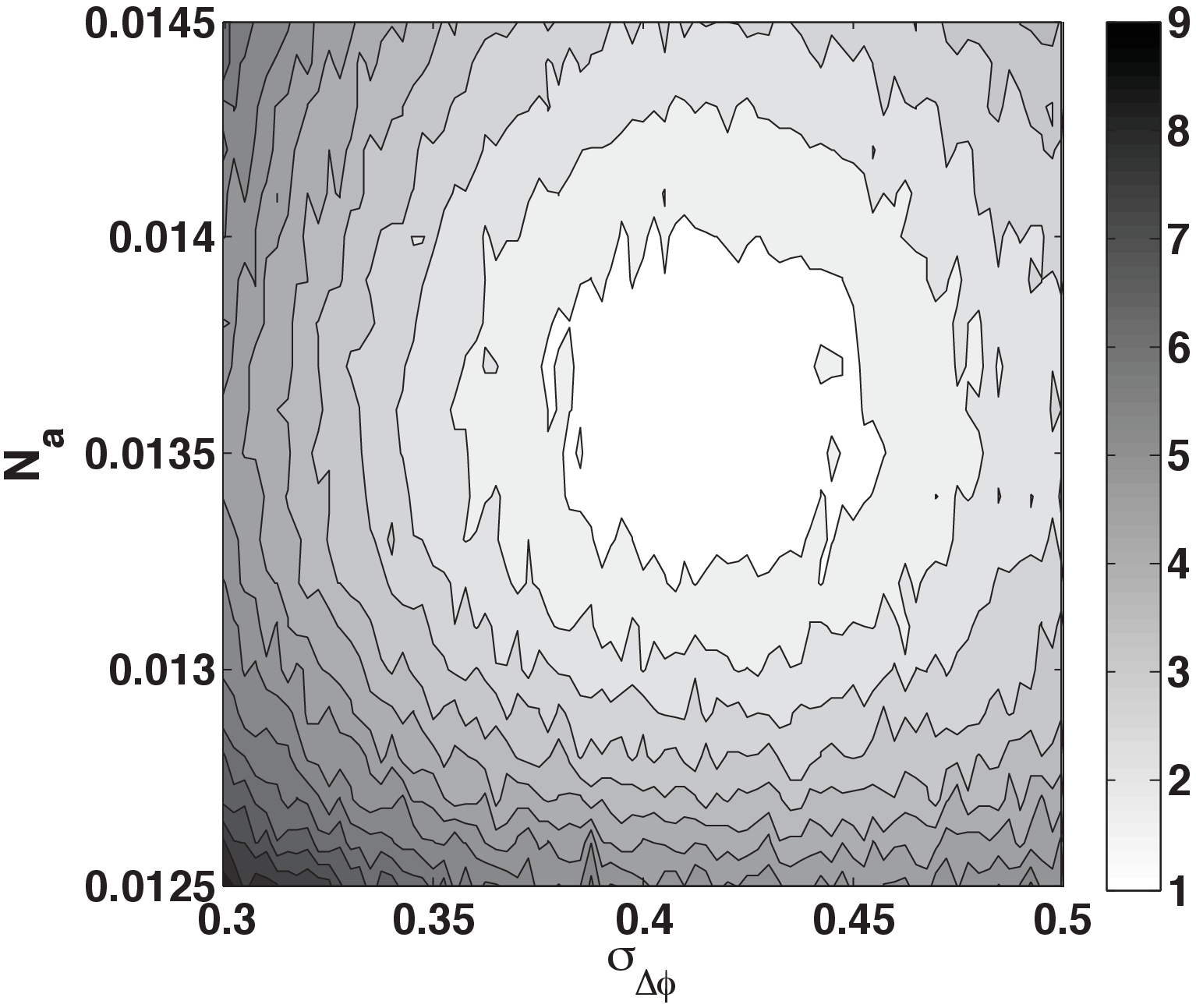} 
\caption{Left: variation of the reduced $\chi^2$ -- measuring the goodness of the fit to the observed data -- as a function of the astrophysical null depth $N_{a}$. The mean and standard deviation of the phase error are left as free parameters, and adjusted to minimize the $\chi^{2}$ for each new value of $N_a$. Center: projected $\chi^2$ map of our model in the $N_a$ vs $\mu_{\Delta\phi}$ plane. For each point, $\sigma_{\Delta\phi}$ is chosen to minimize the $\chi^2$. Right: same map but projected in the $N_a$ vs  $\sigma_{\Delta\phi}$ plane. For these two maps, the contours are over-plotted for each $\chi^2$ intervals of 0.5.}
\label{fig:covariance}
\end{figure}

In section \ref{section_fit}, we described our fitting strategies, and developed the minimization process used to fit the distribution corresponding to an individual null sequence. The statistical error bar $\sigma_{\rm stat}$ on the derived astrophysical null depth is then determined by applying small fluctuations to $N_a$ around its best fit value. For every new value of $N_a$, the two other parameters ($\mu_{\Delta\phi}$ and $\sigma_{\Delta\phi}$) are adjusted to minimize the $\chi^2$. As the number of degrees of freedom of our system is known and is $N_{\rm bins}-4$, it is possible to calculate the $\Delta\chi^2$ relative to a certain confidence level. The error bars are generally evaluated for $1\sigma$ confidence levels, and so we use this criteria. $\sigma_{\rm stat}$ corresponds to the increment in $N_a$ required to increase the reduced $\chi^2$ from its minimum value $\chi^{2}_{\rm min}$ to $\chi^{2}_{\rm min}+\Delta\chi^2$. For the dataset obtained on $\alpha$\,Boo with the PFN, the $1\sigma$ error bar corresponds to a very small $\chi^2$ increment, $\Delta\chi^2=0.07$ and we find $\sigma_{\rm stat}= 3\times10^{-4}$ for the NSC (see Fig.\ref{fig:covariance}, left). Another way of calculating the error bars consist in using bootstrapping methods. We double-checked our confidence interval using this technique and found similar error bars ($\simeq 3\times10^{-4}$). This error bar takes into account the fitting noise that is not present for the ASC.
\\
The central and right panels of Fig. \ref{fig:covariance} represent the normalized $\chi^2$ of our fits projected in two different parameters planes (i.e. $N_a$ vs. $\mu_{\Delta\phi}$ for the central panel and $N_a$ vs. $\sigma_{\Delta\phi}$ for the right one). The contours on these maps are displayed for increments of the $\chi^2$ of $\Delta\chi^2=0.5$. These maps illustrate the covariance of the fits with the two free parameters: the mean and standard deviation of the phase error fluctuations. They show that relatively large variations on the fitted values of these phase parameters, between 0.05 and 0.1 rad, only produce a marginal effect on the measured astrophysical null, smaller than $10^{-3}$, but produce very large effects on the fit quality. This result is  important as it clearly illustrates the resilience of our approach to possible error on the assessment of the phase fluctuations.

\bibliographystyle{apj}	
\bibliography{aeos_bib}

\begin{thebibliography}{26}
\expandafter\ifx\csname natexlab\endcsname\relax\def\natexlab#1{#1}\fi

\bibitem[{{Absil} {et~al.}(2006){Absil}, {den Hartog}, {Gondoin}, {Fabry},
  {Wilhelm}, {Gitton}, \& {Puech}}]{Absil06a}
{Absil}, O., {den Hartog}, R., {Gondoin}, P., {Fabry}, P., {Wilhelm}, R.,
  {Gitton}, P., \& {Puech}, F. 2006, \aap, 448, 787

\bibitem[{{Absil} {et~al.}(2011){Absil}, {den Hartog}, {Gondoin}, {Fabry},
  {Wilhelm}, {Gitton}, \& {Puech}}]{Absil11}
---. 2011, \aap, In press

\bibitem[{{Bracewell}(1978)}]{Bracewell78}
{Bracewell}, R.~N. 1978, \nat, 274, 780

\bibitem[{Cagigal \& Canales(2000)}]{Cagigal00}
Cagigal, M.~P., \& Canales, V.~F. 2000, J. Opt. Soc. Am. A, 17, 1312

\bibitem[{{Claret} {et~al.}(1995){Claret}, {Diaz-Cordoves}, \&
  {Gimenez}}]{Claret95}
{Claret}, A., {Diaz-Cordoves}, J., \& {Gimenez}, A. 1995, \aaps, 114, 247

\bibitem[{{Cochran}(1954)}]{Cochran54}
{Cochran}, W.~G. 1954, Biometrics, 10, 417

\bibitem[{{Colavita} {et~al.}(2009){Colavita}, {Serabyn}, {Millan-Gabet},
  {Koresko}, {Akeson}, {Booth}, {Mennesson}, {Ragland}, {Appleby}, {Berkey},
  {Cooper}, {Crawford}, {Creech-Eakman}, {Dahl}, {Felizardo},
  {Garcia-Gathright}, {Gathright}, {Herstein}, {Hovland}, {Hrynevych}, {Ligon},
  {Medeiros}, {Moore}, {Morrison}, {Paine}, {Palmer}, {Panteleeva}, {Smith},
  {Swain}, {Smythe}, {Summers}, {Tsubota}, {Tyau}, {Vasisht}, {Wetherell},
  {Wizinowich}, \& {Woillez}}]{Colavita09}
{Colavita}, M.~M., {Serabyn}, E., {Millan-Gabet}, R., {Koresko}, C.~D.,
  {Akeson}, R.~L., {Booth}, A.~J., {Mennesson}, B.~P., {Ragland}, S.~D.,
  {Appleby}, E.~C., {Berkey}, B.~C., {Cooper}, A., {Crawford}, S.~L.,
  {Creech-Eakman}, M.~J., {Dahl}, W., {Felizardo}, C., {Garcia-Gathright},
  J.~I., {Gathright}, J.~T., {Herstein}, J.~S., {Hovland}, E.~E., {Hrynevych},
  M.~A., {Ligon}, E.~R., {Medeiros}, D.~W., {Moore}, J.~D., {Morrison}, D.,
  {Paine}, C.~G., {Palmer}, D.~L., {Panteleeva}, T., {Smith}, B., {Swain},
  M.~R., {Smythe}, R.~F., {Summers}, K.~R., {Tsubota}, K., {Tyau}, C.,
  {Vasisht}, G., {Wetherell}, E., {Wizinowich}, P.~L., \& {Woillez}, J.~M.
  2009, \pasp, 121, 1120

\bibitem[{{di Benedetto} \& {Rabbia}(1987)}]{diBenedetto87}
{di Benedetto}, G.~P., \& {Rabbia}, Y. 1987, \aap, 188, 114

\bibitem[{{Haguenauer} \& {Serabyn}(2006)}]{Haguenauer06}
{Haguenauer}, P., \& {Serabyn}, E. 2006, Appl. Opt., 45, 2749

\bibitem[{{Hinz} {et~al.}(2000){Hinz}, {Angel}, {Woolf}, {Hoffmann}, \&
  {McCarthy}}]{Hinz00}
{Hinz}, P.~M., {Angel}, J.~R.~P., {Woolf}, N.~J., {Hoffmann}, W.~F., \&
  {McCarthy}, D.~W. 2000, in Proc. SPIE, Vol. 4006, Interferometry in Optical
  Astronomy, ed. P.~{L\'ena} \& A.~{Quirrenbach}, 349--353

\bibitem[{{Kervella} {et~al.}(2004){Kervella}, {Coud{\'e} du Foresto},
  {Segransan}, \& {di Folco}}]{Kervella04}
{Kervella}, P., {Coud{\'e} du Foresto}, V., {Segransan}, D., \& {di Folco}, E.
  2004, in Society of Photo-Optical Instrumentation Engineers (SPIE) Conference
  Series, Vol. 5491, Society of Photo-Optical Instrumentation Engineers (SPIE)
  Conference Series, ed. {W.~A.~Traub}, 741--+

\bibitem[{{Lacour} {et~al.}(2008){Lacour}, {Meimon}, {Thi{\'e}baut}, {Perrin},
  {Verhoelst}, {Pedretti}, {Schuller}, {Mugnier}, {Monnier}, {Berger},
  {Haubois}, {Poncelet}, {Le Besnerais}, {Eriksson}, {Millan-Gabet}, {Ragland},
  {Lacasse}, \& {Traub}}]{Lacour08}
{Lacour}, S., {Meimon}, S., {Thi{\'e}baut}, E., {Perrin}, G., {Verhoelst}, T.,
  {Pedretti}, E., {Schuller}, P.~A., {Mugnier}, L., {Monnier}, J., {Berger},
  J.~P., {Haubois}, X., {Poncelet}, A., {Le Besnerais}, G., {Eriksson}, K.,
  {Millan-Gabet}, R., {Ragland}, S., {Lacasse}, M., \& {Traub}, W. 2008, \aap,
  485, 561

\bibitem[{{Lay}(2004)}]{Lay04}
{Lay}, O. 2004, \ao, 43, 6100

\bibitem[{{Martin} {et~al.}(2008){Martin}, {Serabyn}, {Liewer}, {Loya},
  {Mennesson}, {Hanot}, \& {Mawet}}]{Martin08}
{Martin}, S., {Serabyn}, E., {Liewer}, K., {Loya}, F., {Mennesson}, B.,
  {Hanot}, C., \& {Mawet}, D. 2008, in Society of Photo-Optical Instrumentation
  Engineers (SPIE) Conference Series, Vol. 7013, Society of Photo-Optical
  Instrumentation Engineers (SPIE) Conference Series

\bibitem[{{Mayor} \& {Queloz}(1995)}]{Mayor95}
{Mayor}, M., \& {Queloz}, D. 1995, \nat, 378, 355

\bibitem[{{Mennesson} {et~al.}(2006){Mennesson}, {Haguenauer}, {Serabyn}, \&
  {Liewer}}]{Mennesson06}
{Mennesson}, B., {Haguenauer}, P., {Serabyn}, E., \& {Liewer}, K. 2006, in
  Society of Photo-Optical Instrumentation Engineers (SPIE) Conference Series,
  Vol. 6268, Society of Photo-Optical Instrumentation Engineers (SPIE)
  Conference Series

\bibitem[{{M{\'e}rand} {et~al.}(2005){M{\'e}rand}, {Bord{\'e}}, \& {Coud{\'e}
  Du Foresto}}]{Merand05}
{M{\'e}rand}, A., {Bord{\'e}}, P., \& {Coud{\'e} Du Foresto}, V. 2005, \aap,
  433, 1155

\bibitem[{{Perrin} {et~al.}(1998){Perrin}, {Coude Du Foresto}, {Ridgway},
  {Mariotti}, {Traub}, {Carleton}, \& {Lacasse}}]{Perrin98}
{Perrin}, G., {Coude Du Foresto}, V., {Ridgway}, S.~T., {Mariotti}, J.,
  {Traub}, W.~A., {Carleton}, N.~P., \& {Lacasse}, M.~G. 1998, in Astronomical
  Society of the Pacific Conference Series, Vol. 154, Cool Stars, Stellar
  Systems, and the Sun, ed. {R.~A.~Donahue \& J.~A.~Bookbinder}, 2021--+

\bibitem[{{Press} {et~al.}(2007){Press}, {Teukolsky}, {Vetterling}, \&
  {Flannery}}]{num-recip}
{Press}, W., {Teukolsky}, S., {Vetterling}, W., \& {Flannery}, B. 2007,
  Numerical Recipes 3rd Edition: The Art of Scientific Computing (Cambridge
  University Press)

\bibitem[{{Rayson} {et~al.}(2004){Rayson}, {Berridge}, \& {Francis}}]{Rayson04}
{Rayson}, P., {Berridge}, D., \& {Francis}, B. 2004, in  (Louvain, Belgium:
  Presses universitaires de Louvain)

\bibitem[{{Riaud} \& {Hanot}(2010)}]{Riaud10}
{Riaud}, P., \& {Hanot}, C. 2010, \apj, 719, 749

\bibitem[{{Roddier}(1983)}]{Roddier83}
{Roddier}, F. 1983, in European Southern Observatory Astrophysics Symposia,
  Vol.~17, European Southern Observatory Astrophysics Symposia, ed.
  {J.-P.~Swings \& K.~Kjaer}, 255--261

\bibitem[{Rohatgi(1976)}]{Rohatgi76}
Rohatgi, V. 1976, An Introduction to Probability Theory Mathematical Statistics
  (Wiley, New York), 141

\bibitem[{{Serabyn}(2000)}]{Serabyn00}
{Serabyn}, E. 2000, in Proc. SPIE, Vol. 4006, Interferometry in Optical
  Astronomy, ed. P.~{L\'ena} \& A.~{Quirrenbach}, 328--339

\bibitem[{{Serabyn} \& {Mennesson}(2006)}]{Serabyn06a}
{Serabyn}, E., \& {Mennesson}, B. 2006, in IAU Colloq. 200: Direct Imaging of
  Exoplanets: Science and Techniques, ed. {C.~Aime \& F.~Vakili}, 379--384

\bibitem[{{Serabyn} {et~al.}(2006){Serabyn}, {Wallace}, {Troy}, {Mennesson},
  {Haguenauer}, {Gappinger}, \& {Bloemhof}}]{Serabyn06b}
{Serabyn}, E., {Wallace}, J.~K., {Troy}, M., {Mennesson}, B., {Haguenauer}, P.,
  {Gappinger}, R.~O., \& {Bloemhof}, E.~E. 2006, in Society of Photo-Optical
  Instrumentation Engineers (SPIE) Conference Series, Vol. 6272, Society of
  Photo-Optical Instrumentation Engineers (SPIE) Conference Series

\end{thebibliography}

\end{document}